\documentclass[11pt]{article}
\usepackage{axodraw}
\usepackage{epsfig}
\usepackage{amsfonts}
\usepackage{amsmath}
\usepackage{bm,bbm}
\usepackage{cite}
\usepackage{hyperref}
\allowdisplaybreaks[1]

\input pix.sty

\renewcommand{\eq}{eq.~}
\renewcommand{\eqs}{eqs.~}
\renewcommand{\se}{sec.~}
\renewcommand{\ses}{secs.~}
\renewcommand{\fig}{fig.~}

\newcommand{\PT}{\mathbbm{P}^\rmii{T}}
\newcommand{\PE}{\mathbbm{P}^\rmii{E}}
\newcommand{\aL}{a^{ }_\rmii{L}}

\newcommand{\E}{\rmii{$E$}}

\newcommand{\Nc}{N_{\rm c}}

\newcommand{\nG}{n_\rmii{G}}
\newcommand{\nS}{n_\rmii{S}}

\newcommand{\mE}{m_\rmii{E}}
\newcommand{\mEtiny}{m_\rmiii{E}}
\newcommand{\mA}{m_\rmii{A}}

\newcommand{\qm}{q_-}
\newcommand{\qp}{q_+}
\newcommand{\qmp}{q_\mp}
\newcommand{\qpm}{q_\pm}

\newcommand{\rmO}{{\mathcal{O}}}

\newcommand{\CF}{C_\rmii{F}}

\def\lsi{\raise0.3ex\hbox{$<$\kern-0.75em\raise-1.1ex\hbox{$\sim$}}}
\def\gsi{\raise0.3ex\hbox{$>$\kern-0.75em\raise-1.1ex\hbox{$\sim$}}}
\newcommand{\lsim}{\mathop{\lsi}}

\newcommand{\nF}{n_\rmii{F}}
\newcommand{\nB}{n_\rmii{B}}
\newcommand{\rmii}[1]{{\mbox{\tiny\rm{#1}}}}
\newcommand{\rmiii}[1]{{\mbox{\tiny{$\scriptstyle{\rm#1}$}}}}

\newcommand{\im}{\mathop{\mbox{Im}}}

\newcommand{\Tint}[1]{{\hbox{$\sum$}\!\!\!\!\!\!\!\int\,}_{\!\!\!\!\raise-0.9ex\hbox{$\scriptstyle{#1}$}}}
\newcommand{\Tinti}[1]{{{\Sigma}\!\!\!\!\raise0.3ex\hbox{$\int$}_\rmii{${#1}$}}}

\newcommand{\ZZ}{{\mathbb{Z}}}

\newcommand{\bi}{\begin{itemize}}
\newcommand{\ei}{\end{itemize}}
\newcommand{\hide}[1]{ }

\def\TAsc(#1,#2)(#3,#4,#5)%
{\SetWidth{2.0}\CArc(#1,#2)(#3,#4,#5)\SetWidth{1.0}}
\def\Lwidth{3}

\def\TAgl(#1,#2)(#3,#4,#5){\SetWidth{2.0}\PhotonArc(#1,#2)(#3,#4,#5){\Lwidth}%
{6.283 #3 mul 360 div #4 #5 sub #4 #5 sub mul sqrt mul Tdensity mul}%
\SetWidth{1.0}}
\def\TLgl(#1,#2)(#3,#4){\SetWidth{2.0}\Photon(#1,#2)(#3,#4){\Lwidth}
{#1 #3 sub #1 #3 sub mul #2 #4 sub #2 #4 sub mul add sqrt Tdensity mul}%
\SetWidth{1.0}}

\renewcommand{\picc}[1]{\;\parbox[c]{60pt}{\begin{picture}(60,30)(0,-10)
\SetWidth{1.0}\SetScale{1.3} #1 \end{picture}}\; }
\def\Lwidth{1.3}

\newcommand{\picu}[1]{\;\parbox[c]{60pt}{\begin{picture}(60,30)(0,-5)
\SetWidth{1.0}\SetScale{0.7} #1 \end{picture}}\; }
\newcommand{\picv}[1]{\;\parbox[c]{60pt}{\begin{picture}(60,30)(0,0)
\SetWidth{1.0}\SetScale{0.7} #1 \end{picture}}\; }
\newcommand{\picg}[1]{\;\parbox[c]{62pt}{\begin{picture}(160,80)(0,-27)
\SetWidth{1.0}\SetScale{0.45} #1 \end{picture}}\;}
\def\Phis{\picu{%
 \Asc(30,5)(22.3,27,153)%
 \Asc(30,25)(22.3,207,333)%
 \COval(10,15)(2,2)(0){Black}{Black}%
 \COval(50,15)(2,2)(0){Black}{Black}%
}}
\def\PhissF{\picu{%
 \Asc(30,5)(22.3,27,153)%
 \Asc(30,25)(22.3,207,333)%
 \COval(10,15)(2,2)(0){Black}{Black}%
 \COval(50,15)(2,2)(0){Black}{Black}%
 \Asc(30,20)(7,15,375)
}}
\def\PhissG{\picu{%
 \Asc(20,10)(11.15,27,153)%
 \Asc(20,20)(11.15,207,333)%
 \Asc(40,10)(11.15,27,153)%
 \Asc(40,20)(11.15,207,333)%
 \COval(10,15)(2,2)(0){Black}{Black}%
 \COval(50,15)(2,2)(0){Black}{Black}%
}}
\def\Phif{\picu{%
 \Aqu(30,5)(22.3,27,153)%
 \Aqu(30,25)(22.3,207,333)%
 \COval(10,15)(2,2)(0){Black}{Black}%
 \COval(50,15)(2,2)(0){Black}{Black}%
}}
\def\Phig{\picu{%
 \Agl(30,5)(22.3,27,153)%
 \Agl(30,25)(22.3,207,333)%
 \COval(10,15)(2,2)(0){Black}{Black}%
 \COval(50,15)(2,2)(0){Black}{Black}%
}}
\def\PhiggA{\picu{%
 \Agl(20,5)(22.3,27,153)%
 \Agl(20,25)(22.3,207,333)%
 \COval(0,15)(2,2)(0){Black}{Black}%
 \COval(40,15)(2,2)(0){Black}{Black}%
 \Agl(48,15)(8,-180,180)%
}}
\def\PhiggB{\picu{%
 \Agl(30,5)(22.3,27,153)%
 \Agl(30,25)(22.3,207,333)%
 \COval(10,15)(2,2)(0){Black}{Black}%
 \COval(50,15)(2,2)(0){Black}{Black}%
 \Lgl(10,15)(50,15)%
}}
\def\PhiggC{\picu{%
 \Agl(30,5)(22.3,90,153)%
 \Agl(30,25)(22.3,207,333)%
 \COval(10,15)(2,2)(0){Black}{Black}%
 \COval(50,15)(2,2)(0){Black}{Black}%
 \Agl(43,27)(12,180,300)%
 \Agl(37,16)(12.5,0,120)%
}}
\def\PhiggD{\picu{%
 \Agl(30,5)(22.3,27,72)%
 \Agl(30,5)(22.3,108,153)%
 \Agl(30,25)(22.3,207,333)%
 \COval(10,15)(2,2)(0){Black}{Black}%
 \COval(50,15)(2,2)(0){Black}{Black}%
 \Agl(30,27.3)(7,0,360)
}}
\def\PhiggE{\picu{%
 \Agl(30,5)(22.3,27,72)%
 \Agl(30,5)(22.3,108,153)%
 \Agl(30,25)(22.3,207,333)%
 \COval(10,15)(2,2)(0){Black}{Black}%
 \COval(50,15)(2,2)(0){Black}{Black}%
 \Agh(30,27.3)(7,0,180)
 \Agh(30,27.3)(7,180,360)
}}
\def\PhiggF{\picu{%
 \Agl(30,5)(22.3,27,153)%
 \Agl(30,25)(22.3,207,333)%
 \COval(10,15)(2,2)(0){Black}{Black}%
 \COval(50,15)(2,2)(0){Black}{Black}%
 \Agl(30,20)(7,15,375)
}}
\def\PhiggG{\picu{%
 \Agl(20,10)(11.15,27,153)%
 \Agl(20,20)(11.15,207,333)%
 \Agl(40,10)(11.15,27,153)%
 \Agl(40,20)(11.15,207,333)%
 \COval(10,15)(2,2)(0){Black}{Black}%
 \COval(50,15)(2,2)(0){Black}{Black}%
}}
\def\PhiggH{\picu{%
 \Agl(30,5)(22.3,27,153)%
 \Agl(30,25)(22.3,207,333)%
 \COval(10,15)(2,2)(0){Black}{Black}%
 \COval(50,15)(2,2)(0){Black}{Black}%
 \Lgl(30,2.7)(30,27.3)%
}}
\def\Phisf{\picu{%
 \Asc(30,5)(22.3,27,72)%
 \Asc(30,5)(22.3,108,153)%
 \Asc(30,25)(22.3,207,333)%
 \COval(10,15)(2,2)(0){Black}{Black}%
 \COval(50,15)(2,2)(0){Black}{Black}%
 \Aqu(30,27.3)(7,0,180)%
 \Aqu(30,27.3)(7,180,360)%
}}
\def\PhifsD{\picu{%
 \Aqu(30,5)(22.3,27,72)%
 \Aqu(30,5)(22.3,108,153)%
 \Aqu(30,25)(22.3,207,333)%
 \COval(10,15)(2,2)(0){Black}{Black}%
 \COval(50,15)(2,2)(0){Black}{Black}%
 \Aqu(30,27.3)(7,0,180)%
 \Asc(30,27.3)(7,180,360)%
}}
\def\PhifsH{\picu{%
 \Aqu(30,5)(22.3,27,90)%
 \Aqu(30,5)(22.3,90,153)%
 \Aqu(30,25)(22.3,207,270)%
 \Aqu(30,25)(22.3,270,333)%
 \COval(10,15)(2,2)(0){Black}{Black}%
 \COval(50,15)(2,2)(0){Black}{Black}%
 \Lsc(30,2.7)(30,27.3)%
}}
\def\Phisxf{\picu{%
 \Aqu(30,5)(22.3,27,90)%
 \Asc(30,5)(22.3,90,153)%
 \Asc(30,25)(22.3,207,270)%
 \Aqu(30,25)(22.3,270,333)%
 \COval(10,15)(2,2)(0){Black}{Black}%
 \COval(50,15)(2,2)(0){Black}{Black}%
 \Laqu(30,2.7)(30,27.3)%
}}
\def\PhigsD{\picu{%
 \Agl(30,5)(22.3,27,72)%
 \Agl(30,5)(22.3,108,153)%
 \Agl(30,25)(22.3,207,333)%
 \COval(10,15)(2,2)(0){Black}{Black}%
 \COval(50,15)(2,2)(0){Black}{Black}%
 \Asc(30,27.3)(7,0,180)%
 \Asc(30,27.3)(7,180,360)%
}}
\def\PhigsF{\picu{%
 \Agl(30,5)(22.3,27,153)%
 \Agl(30,25)(22.3,207,333)%
 \COval(10,15)(2,2)(0){Black}{Black}%
 \COval(50,15)(2,2)(0){Black}{Black}%
 \Asc(30,20)(7,15,375)
}}
\def\PhisgA{\picu{%
 \Asc(20,5)(22.3,27,153)%
 \Asc(20,25)(22.3,207,333)%
 \COval(0,15)(2,2)(0){Black}{Black}%
 \COval(40,15)(2,2)(0){Black}{Black}%
 \Agl(48,15)(8,-180,180)%
}}
\def\PhisgB{\picu{%
 \Asc(30,5)(22.3,27,153)%
 \Asc(30,25)(22.3,207,333)%
 \COval(10,15)(2,2)(0){Black}{Black}%
 \COval(50,15)(2,2)(0){Black}{Black}%
 \Lgl(10,15)(50,15)%
}}
\def\PhisgC{\picu{%
 \Asc(30,5)(22.3,88,153)%
 \Asc(30,25)(22.3,207,333)%
 \COval(10,15)(2,2)(0){Black}{Black}%
 \COval(50,15)(2,2)(0){Black}{Black}%
 \Agl(43,27)(12,180,300)%
 \Asc(37,16)(12.5,0,120)%
}}
\def\PhisgD{\picu{%
 \Asc(30,5)(22.3,27,72)%
 \Asc(30,5)(22.3,108,153)%
 \Asc(30,25)(22.3,207,333)%
 \COval(10,15)(2,2)(0){Black}{Black}%
 \COval(50,15)(2,2)(0){Black}{Black}%
 \Asc(30,27.3)(7,0,180)
 \Agl(30,27.3)(7,180,360)
}}
\def\PhisgF{\picu{%
 \Asc(30,5)(22.3,27,153)%
 \Asc(30,25)(22.3,207,333)%
 \COval(10,15)(2,2)(0){Black}{Black}%
 \COval(50,15)(2,2)(0){Black}{Black}%
 \Agl(30,20)(7,15,375)
}}
\def\PhisgH{\picu{%
 \Asc(30,5)(22.3,27,153)%
 \Asc(30,25)(22.3,207,333)%
 \COval(10,15)(2,2)(0){Black}{Black}%
 \COval(50,15)(2,2)(0){Black}{Black}%
 \Lgl(30,2.7)(30,27.3)%
}}
\def\PhisxgA{\picu{%
 \Agl(40,5)(22.3,27,153)%
 \Agl(40,25)(22.3,207,333)%
 \COval(20,15)(2,2)(0){Black}{Black}%
 \COval(60,15)(2,2)(0){Black}{Black}%
 \Asc(12,15)(8,0,360)%
}}
\def\PhisxgC{\picu{%
 \Agl(30,5)(22.3,27,90)%
 \Agl(30,25)(22.3,207,333)%
 \COval(10,15)(2,2)(0){Black}{Black}%
 \COval(50,15)(2,2)(0){Black}{Black}%
 \Asc(17,27)(12,240,360)%
 \Asc(22,16)(12,60,180)%
}}
\def\PhisxgG{\picu{%
 \Asc(20,10)(11.15,27,153)%
 \Asc(20,20)(11.15,207,333)%
 \Agl(40,10)(11.15,27,153)%
 \Agl(40,20)(11.15,207,333)%
 \COval(10,15)(2,2)(0){Black}{Black}%
 \COval(50,15)(2,2)(0){Black}{Black}%
}}
\def\PhisxgH{\picu{%
 \Agl(30,5)(22.3,27,90)%
 \Asc(30,5)(22.3,90,153)%
 \Asc(30,25)(22.3,207,270)%
 \Agl(30,25)(22.3,270,333)%
 \COval(10,15)(2,2)(0){Black}{Black}%
 \COval(50,15)(2,2)(0){Black}{Black}%
 \Lsc(30,2.7)(30,27.3)%
}}
\def\PhigfD{\picu{%
 \Agl(30,5)(22.3,27,72)%
 \Agl(30,5)(22.3,108,153)%
 \Agl(30,25)(22.3,207,333)%
 \COval(10,15)(2,2)(0){Black}{Black}%
 \COval(50,15)(2,2)(0){Black}{Black}%
 \Aqu(30,27.3)(7,0,180)%
 \Aqu(30,27.3)(7,180,360)%
}}
\def\PhifgB{\picu{%
 \Aqu(30,5)(22.3,27,153)%
 \Aqu(30,25)(22.3,207,333)%
 \COval(10,15)(2,2)(0){Black}{Black}%
 \COval(50,15)(2,2)(0){Black}{Black}%
 \Lgl(10,15)(50,15)%
}}
\def\PhifgC{\picu{%
 \Aqu(30,5)(22.3,88,153)%
 \Aqu(30,25)(22.3,207,333)%
 \COval(10,15)(2,2)(0){Black}{Black}%
 \COval(50,15)(2,2)(0){Black}{Black}%
 \Agl(43,27)(12,180,300)%
 \Aqu(37,16)(12.5,0,120)%
}}
\def\PhifgD{\picu{%
 \Aqu(30,5)(22.3,27,72)%
 \Aqu(30,5)(22.3,108,153)%
 \Aqu(30,25)(22.3,207,333)%
 \COval(10,15)(2,2)(0){Black}{Black}%
 \COval(50,15)(2,2)(0){Black}{Black}%
 \Aqu(30,27.3)(7,0,180)
 \Agl(30,27.3)(7,180,360)
}}
\def\PhifgH{\picu{%
 \Aqu(30,5)(22.3,27,90)%
 \Aqu(30,5)(22.3,90,153)%
 \Aqu(30,25)(22.3,207,270)%
 \Aqu(30,25)(22.3,270,333)%
 \COval(10,15)(2,2)(0){Black}{Black}%
 \COval(50,15)(2,2)(0){Black}{Black}%
 \Lgl(30,2.7)(30,27.3)%
}}
\def\PhifxgC{\picu{%
 \Agl(30,5)(22.3,27,92)%
 \Agl(30,25)(22.3,207,333)%
 \COval(10,15)(2,2)(0){Black}{Black}%
 \COval(50,15)(2,2)(0){Black}{Black}%
 \Aqu(17,27)(12.5,240,360)%
 \Aqu(23,16)(12.5,60,180)%
}}
\def\PhifxgH{\picu{%
 \Agl(30,5)(22.3,27,90)%
 \Aqu(30,5)(22.3,90,153)%
 \Aqu(30,25)(22.3,207,270)%
 \Agl(30,25)(22.3,270,333)%
 \COval(10,15)(2,2)(0){Black}{Black}%
 \COval(50,15)(2,2)(0){Black}{Black}%
 \Lqu(30,2.7)(30,27.3)%
}}
\def\Cut{\picc{%
 \Asc(30,5)(22.3,27,90)%
 \Asc(30,5)(22.3,90,153)%
 \Asc(30,25)(22.3,207,270)%
 \Asc(30,25)(22.3,270,333)%
 \COval(10,15)(2,2)(0){Black}{Black}%
 \COval(50,15)(2,2)(0){Black}{Black}%
 \Lsc(30,2.7)(30,27.3)%
 \SetWidth{1.0}%
 \Line(10,25)(50,5)
 \Line(38.5,23)(44.5,25)
 \Line(40.5,27)(42.5,21)
 \Line(21.5,7)(15.5,5)
 \Line(19.5,3)(17.5,9)
 \Text(58,38)[c]{$\scriptstyle b$}
 \Text(20,2)[c]{$\scriptstyle d$}
 \Text(25,38)[c]{$\scriptstyle \sigma_a$}
 \Text(54,1)[c]{$\scriptstyle \sigma_e$}
 \Text(45,25)[c]{$\scriptstyle \sigma_c$}
}}
 \def\GraphggA{\picv{
  \SetWidth{1.0}
  \Lgl(0,5)(25,10)
  \Lgl(25,10)(50,5)
  \Lgl(25,10)(25,30)
  \Lgl(0,37.5)(25,32.5)
  \Lgr(25.5,30)(50.5,35)
  \Lgr(24.9,32.9)(49.9,37.9)
  \GCirc(25,31){2}{0}
 }}
 \def\GraphggB{\picv{
  \SetWidth{1.0}
  \Lgl(0,5)(25,20)
  \Lgl(25,20)(50,5)
  \Lgl(0,37.5)(25,22.5)
  \Lgr(26.4,20.0)(51.4,35.0)
  \Lgr(25,22.5)(50,37.5)
  \GCirc(25,21.25){2}{0}
 }}
 \def\GraphsfA{\picv{
  \SetWidth{1.0}
  \Lqu(0,5)(25,10)
  \Lqu(25,10)(50,5)
  \Lsc(25,10)(25,30)
  \Lsc(0,37.5)(25,32.5)
  \Lgr(25.5,30)(50.5,35)
  \Lgr(24.9,32.9)(49.9,37.9)
  \GCirc(25,31){2}{0}
 }}
 \def\GraphsfB{\picv{
  \SetWidth{1.0}
  \Laqu(0,5)(25,10)
  \Laqu(25,10)(50,5)
  \Lsc(25,10)(25,30)
  \Lsc(0,37.5)(25,32.5)
  \Lgr(25.5,30)(50.5,35)
  \Lgr(24.9,32.9)(49.9,37.9)
  \GCirc(25,31){2}{0}
 }}
 \def\GraphsfC{\picv{
  \SetWidth{1.0}
  \Lsc(0,5)(25,10)
  \Laqu(25,10)(50,5)
  \Lqu(25,10)(25,30)
  \Laqu(0,37.5)(25,32.5)
  \Lgr(25.5,30)(50.5,35)
  \Lgr(24.9,32.9)(49.9,37.9)
  \GCirc(25,31){2}{0}
 }}
 \def\GraphsfD{\picv{
  \SetWidth{1.0}
  \Lqu(0,5)(25,10)
  \Lsc(25,10)(50,5)
  \Lqu(25,10)(25,30)
  \Laqu(0,37.5)(25,32.5)
  \Lgr(25.5,30)(50.5,35)
  \Lgr(24.9,32.9)(49.9,37.9)
  \GCirc(25,31){2}{0}
 }}
 \def\GraphsfE{\picv{
  \SetWidth{1.0}
  \Lsc(0,5)(25,10)
  \Lqu(25,10)(50,5)
  \Laqu(25,10)(25,30)
  \Lqu(0,37.5)(25,32.5)
  \Lgr(25.5,30)(50.5,35)
  \Lgr(24.9,32.9)(49.9,37.9)
  \GCirc(25,31){2}{0}
 }}
 \def\GraphsfF{\picv{
  \SetWidth{1.0}
  \Laqu(0,5)(25,10)
  \Lsc(25,10)(50,5)
  \Laqu(25,10)(25,30)
  \Lqu(0,37.5)(25,32.5)
  \Lgr(25.5,30)(50.5,35)
  \Lgr(24.9,32.9)(49.9,37.9)
  \GCirc(25,31){2}{0}
 }}
 \def\GraphsgA{\picv{
  \SetWidth{1.0}
  \Lsc(0,5)(25,10)
  \Lsc(25,10)(50,5)
  \Lgl(25,10)(25,30)
  \Lgl(0,37.5)(25,32.5)
  \Lgr(25.5,30)(50.5,35)
  \Lgr(24.9,32.9)(49.9,37.9)
  \GCirc(25,31){2}{0}
 }}
 \def\GraphsgB{\picv{
  \SetWidth{1.0}
  \Lgl(0,5)(25,10)
  \Lsc(25,10)(50,5)
  \Lsc(25,10)(25,30)
  \Lsc(0,37.5)(25,32.5)
  \Lgr(25.5,30)(50.5,35)
  \Lgr(24.9,32.9)(49.9,37.9)
  \GCirc(25,31){2}{0}
 }}
 \def\GraphsgC{\picv{
  \SetWidth{1.0}
  \Lsc(0,5)(25,10)
  \Lgl(25,10)(50,5)
  \Lsc(25,10)(25,30)
  \Lsc(0,37.5)(25,32.5)
  \Lgr(25.5,30)(50.5,35)
  \Lgr(24.9,32.9)(49.9,37.9)
  \GCirc(25,31){2}{0}
 }}
 \def\GraphsgD{\picv{
  \SetWidth{1.0}
  \Lsc(0,5)(25,20)
  \Lsc(25,20)(50,5)
  \Lgl(0,37.5)(25,22.5)
  \Lgr(26.4,20.0)(51.4,35.0)
  \Lgr(25,22.5)(50,37.5)
  \GCirc(25,21.25){2}{0}
 }}
 \def\GraphfgA{\picv{
  \SetWidth{1.0}
  \Lqu(0,5)(25,10)
  \Lqu(25,10)(50,5)
  \Lgl(25,10)(25,30)
  \Lgl(0,37.5)(25,32.5)
  \Lgr(25.5,30)(50.5,35)
  \Lgr(24.9,32.9)(49.9,37.9)
  \GCirc(25,31){2}{0}
 }}
 \def\GraphfgB{\picv{
  \SetWidth{1.0}
  \Laqu(0,5)(25,10)
  \Laqu(25,10)(50,5)
  \Lgl(25,10)(25,30)
  \Lgl(0,37.5)(25,32.5)
  \Lgr(25.5,30)(50.5,35)
  \Lgr(24.9,32.9)(49.9,37.9)
  \GCirc(25,31){2}{0}
 }}
 \def\GraphfgC{\picv{
  \SetWidth{1.0}
  \Lgl(0,5)(25,10)
  \Laqu(25,10)(50,5)
  \Lqu(25,10)(25,30)
  \Laqu(0,37.5)(25,32.5)
  \Lgr(25.5,30)(50.5,35)
  \Lgr(24.9,32.9)(49.9,37.9)
  \GCirc(25,31){2}{0}
 }}
 \def\GraphfgD{\picv{
  \SetWidth{1.0}
  \Lqu(0,5)(25,10)
  \Lgl(25,10)(50,5)
  \Lqu(25,10)(25,30)
  \Laqu(0,37.5)(25,32.5)
  \Lgr(25.5,30)(50.5,35)
  \Lgr(24.9,32.9)(49.9,37.9)
  \GCirc(25,31){2}{0}
 }}
 \def\GraphfgE{\picv{
  \SetWidth{1.0}
  \Lgl(0,5)(25,10)
  \Lqu(25,10)(50,5)
  \Laqu(25,10)(25,30)
  \Lqu(0,37.5)(25,32.5)
  \Lgr(25.5,30)(50.5,35)
  \Lgr(24.9,32.9)(49.9,37.9)
  \GCirc(25,31){2}{0}
 }}
 \def\GraphfgF{\picv{
  \SetWidth{1.0}
  \Laqu(0,5)(25,10)
  \Lgl(25,10)(50,5)
  \Laqu(25,10)(25,30)
  \Lqu(0,37.5)(25,32.5)
  \Lgr(25.5,30)(50.5,35)
  \Lgr(24.9,32.9)(49.9,37.9)
  \GCirc(25,31){2}{0}
 }}
 \def\GraphfgH{\picv{
  \SetWidth{1.0}
  \Lqu(0,5)(25,20)
  \Lqu(25,20)(50,5)
  \Lgl(0,37.5)(25,22.5)
  \Lgr(26.4,20.0)(51.4,35.0)
  \Lgr(25,22.5)(50,37.5)
  \GCirc(25,21.25){2}{0}
 }}
 \def\GraphggC{\picg{
  \SetWidth{1.5}
  \Photon(0,10)(50,20){3}{6}
  \Photon(50,20)(100,10){3}{6}
  \Photon(50,20)(50,60){3}{5}
  \Photon(0,70)(50,60){3}{6}
  \Photon(50,60)(100,72.5){3}{6}
  \Photon(100,70)(150,40){3}{6}
  \Photon(103,71)(153,101){1}{6}
  \Photon(100,75)(150,105){1}{6}
  \GCirc(100,72.5){5}{0}
 }}
\makeatletter \@addtoreset{equation}{section} \makeatother
\renewcommand{\theequation}{\arabic{section}.\arabic{equation}}
\makeatletter
\renewcommand\section{\@startsection {section}{1}{\z@}%
                                   {-5.5ex \@plus -1ex \@minus -.2ex}
                                   {2.3ex \@plus.2ex}%
                                   {\normalfont\large\bfseries}}
\renewcommand\subsection{\@startsection{subsection}{2}{\z@}%
                                     {-3.25ex\@plus -1ex \@minus -.2ex}%
                                     {1.5ex \@plus .2ex}%
                                     {\normalfont\normalsize\bfseries}}
\renewcommand\thesection {\@arabic\c@section}
\renewcommand\thesubsection   {\thesection.\@arabic\c@subsection}
\renewcommand{\@seccntformat}[1]{%
\csname the#1\endcsname.\hspace{1.0em}}
\makeatother

\begin{document}

\flushbottom

\begin{titlepage}

\begin{flushright}
June 2020
\end{flushright}
\begin{centering}
\vfill

{\Large{\bf
 Gravitational wave background from Standard Model physics:
  \\[3mm]
 Complete leading order
}} 

\vspace{0.8cm}

J.~Ghiglieri$^\rmi{a}$, 
G.~Jackson$^\rmi{b}$,
M.~Laine$^\rmi{b}$
 and 
Y.~Zhu$^\rmi{c,}$\footnote{%
 Previous address: 
 Physik-Department, TU M\"unchen, James-Franck-Strasse 1, 
 85748 Garching, Germany
 }
 
\vspace{0.8cm}

$^\rmi{a}$%
{\em
SUBATECH, Universit\'e de Nantes, IMT Atlantique, IN2P3/CNRS,\\
4 rue Alfred Kastler, La Chantrerie BP 20722, 44307 Nantes, France \\}

\vspace{0.3cm}

$^\rmi{b}$%
{\em
AEC, 
Institute for Theoretical Physics, 
University of Bern, \\ 
Sidlerstrasse 5, CH-3012 Bern, Switzerland \\}

\vspace{0.3cm}

$^\rmi{c}$%
{\em
Wetzlar, Germany\\} 

\vspace*{0.8cm}

\mbox{\bf Abstract}
 
\end{centering}

\vspace*{0.3cm}
 
\noindent
We compute the production rate of the energy density carried by
gravitational waves emitted by a Standard Model plasma in thermal
equilibrium, consistently to leading order in coupling constants for
momenta $k\sim \pi T$. Summing up the contributions from the full
history of the universe, the highest temperature of the radiation
epoch can be constrained by the so-called $N^{ }_\rmi{eff}$
parameter. The current theoretical uncertainty 
$\Delta N^{ }_\rmi{eff} \le 10^{-3}$ corresponds to 
$T^{ }_\rmi{max} \le 2\times 10^{17}$~GeV. 
In the course of the computation, we show how a subpart
of the production rate can be determined with the help 
of standard packages, even if subsequently
an IR subtraction and thermal resummation need to be implemented.

\vfill


\end{titlepage}

\tableofcontents

%
\section{Introduction}

A neutral plasma with charged constituents, such as the early 
universe before recombination, emits and absorbs photons, because scatterings
between the microscopic constituents amount to changing electromagnetic 
currents. Similarly, a homogeneous plasma can emit and absorb gravitational
waves, because scatterings also imply changing energy and momentum
currents (cf., e.g., ref.~\cite{sw}). 
The emission/absorption rate is suppressed by $1/m^{2}_\rmi{Pl}$ and
therefore tiny for temperatures much below the Planck scale. On the
other hand, the age of the universe (inverse Hubble rate) is 
$\sim m^{ }_\rmi{Pl}$, so that the total energy
density emitted into gravitational radiation 
is only suppressed by $1/m^{ }_\rmi{Pl}$. 
This may motivate a precise computation
of the production rate and its integration
over the history of the universe~\cite{qualitative}. 

In addition to the emission from an equilibrium plasma, 
there are numerous potential 
non-equilibrium sources for gravitational radiation. 
These range from tensor modes produced during inflation~\cite{tensor} to 
a multitude of post-inflationary sources (for a review see, e.g., 
ref.~\cite{rev}). However, 
all of these rely on yet-to-be-established models,
unlike the Standard Model background that we are interested in. 

Restricting for a moment to locally Minkowskian spacetime, 
the rate of change of the polarization-averaged 
phase space distribution  
of gravitons ($f^{ }_\rmii{GW}$) has the form~\cite{db}
\be
 \dot{f}^{ }_\rmii{GW} (t,\vec{k})
 \; = \; 
 \Gamma(k) \, 
 \bigl[
 \nB^{ }(k) - f^{ }_\rmii{GW}(t,\vec{k}) 
 \bigr]
 + \rmO\biggl( \frac{1}{m_\rmi{Pl}^4}\biggr)
 \;, \la{rate_gen}
\ee
where $k \equiv |\vec{k}|$ and $\nB(k) \equiv 1 / (e^{k/T} - 1)$
is the Bose distribution. 
The differential energy density is given by 
$
 {\rm d}e^{ }_\rmii{GW} 
 = 
 2 k\, f^{ }_\rmii{GW} \, 
 \frac{{\rm d}^3\vec{k}}{(2\pi)^3} 
$. 
Adopting a logarithmic scale, 
the production rate of gravitational energy density 
can thus be expressed as
\be 
 \frac{ {\rm d}e^{ }_\rmii{GW} }{{\rm d}t\,{\rm d}\ln k}
 = \frac{k^4 \dot{f}^{ }_\rmii{GW}}{\pi^2}
 \;.
\ee

In the following we are interested in estimating
the rate $\Gamma(k)$ defined by \eq\nr{rate_gen}
in the frequency range in which 
$
 {\rm d}e^{ }_\rmii{GW} 
$
peaks. This range is given by the typical thermal 
scale $k\sim \pi T$~\cite{qualitative}, corresponding after 
red shift to the same microwave range 
at which most CMB photons lie. 
In this frequency range, the gravitational wave abundance is 
expected to be much below equilibrium, $f^{ }_\rmii{GW} \ll \nB^{ }(k)$, 
so that the right-hand side of \eq\nr{rate_gen} evaluates to 
$\Gamma(k)\nB^{ }(k)$. However, 
the same coefficient $\Gamma(k)$ also
governs other phenomena, for instance the damping of 
a gravitational wave as it passes through a thermal plasma, 
if produced by some
astrophysical source before 
(cf.,\ e.g.,\ refs.~\cite{damp1,damp2} for recent works).  

We start by describing in some detail the technical steps of 
the computation, which we have implemented 
in two complementary ways, 
{\it viz.}\ by taking the cut of a retarded
2-point correlator of the energy-momentum tensor
(\ses\ref{ss:setup}--\ref{ss:2to2}), and by considering Boltzmann
equations for graviton production (\se\ref{ss:boltzmann}). 
After phase space integration (\se\ref{ss:phase_space})
and thermal resummation (\se\ref{ss:htl}), 
the result is evaluated numerically
(\se\ref{se:numerics}) 
and embedded in a cosmological environment
(\se\ref{se:cosmo}). 
Conclusions and an outlook are offered in \se\ref{se:concl}. 
Two appendices explain why two classes of 
contributions, frequently considered in the literature, 
are of subleading order for the present observable.

%
\section{Steps of the computation}
\la{se:steps}

%
\subsection{Setup}
\la{ss:setup} 

Assuming that a system is spatially homogeneous 
and stationary on the time scales
observed, and aligning the $z$-axis with the momentum 
($\vec{k} = k\, \vec{e}^{ }_z$), the production rate of the energy density 
carried by gravitational waves can be related to the  
Wightman correlator  
\be
 G^{<}_{12;12}
 \; \equiv \; 
 \int_\mathcal{X} 
 \! 
 e^{ik(t-z)}
 \bigl\langle 
  T^{ }_{12}(0) 
  \, T^{ }_{12}(\mathcal{X})
 \bigr\rangle
 \;, \quad
 \mathcal{X} \equiv (t,\vec{x})
 \;. 
 \la{master}
\ee
Here we work in the medium rest frame, with its four-velocity
taking the form $u = (1,\vec{0})$, in order to permit for a 
simple identification of the energy density. For a general 
frame, spatial indices $(...)^{ }_i$ should be replaced 
with $({g_{i}^{ }}_{ }^{\mu} - u^{ }_i u_{ }^\mu)(...)^{ }_\mu$. 

In equilibrium, $G^{<}_{12;12}$ is related to the imaginary
part of the retarded correlator as 
$
 G^{<}_{12;12} = 2 \nB(k) \im G^\rmii{R}_{12;12}
$. 
In the following we compute a Euclidean correlator
$ 
 G^\rmii{E}_{12;12}
$
as a function of a Euclidean four-momentum 
$K=(k^{ }_n,\vec{k})$, 
from which 
$ 
 G^\rmii{R}_{12;12}
$
is obtained by an analytic continuation, 
$ 
 G^\rmii{R}_{12;12}
 = 
 G^\rmii{E}_{12;12} |^{ }_{k^{ }_n \to -i [k + i 0^+]}
$. 
Here $k^{ }_n = 2\pi n T$, 
with $n \in \ZZ$, is a bosonic Matsubara frequency. 
The rate $\Gamma(k)$ from \eq\nr{rate_gen} is then 
given by~\cite{qualitative} 
\be
 \Gamma(k) = \frac{16\pi \im G^\rmii{R}_{12;12}}{k\, m_\rmi{Pl}^2}
 \;, \la{Gamma_k}
\ee
where $m^{ }_\rmi{Pl} = 1.22091 \times 10^{19}$~GeV is the Planck mass.

We write the correlator in a covariant form as\footnote{%
 A simple way to verify the factor in the denominator is to 
 consider momentum averages in the transverse plane. By rotational
 symmetry, 
 $
  \langle q^{ }_i q^{ }_j q^{ }_k q^{ }_l\rangle
  = A \, 
 (  \delta^{ }_{ij}\delta^{ }_{kl}
  + \delta^{ }_{ik}\delta^{ }_{jl}
  + \delta^{ }_{il}\delta^{ }_{jk} )
 $.
 Therefore 
 a representative of $\langle T^{ }_{12} T^{ }_{12} \rangle$
 evaluates to 
 $
  \langle q_1^2 q_2^2 \rangle = A
 $, 
 whereas 
 $
  L^{ }_{ij;kl} 
  \langle q^{ }_i q^{ }_j q^{ }_k q^{ }_l\rangle
  = 
  A\, D(D-3)
 $. \la{transverse}
 } 
\be
 G^\rmii{E}_{12;12} 
 = 
 \frac{ 
 L^{ }_{\mu\nu;\alpha\beta}\, 
 G^\rmii{E}_{\mu\nu;\alpha\beta}
 }{D(D-3)} 
 \;, \quad
  G^\rmii{E}_{\mu\nu;\alpha\beta} 
  \; \equiv \; 
  \int_X e^{i K\cdot X}
   \bigl\langle 
     T^{ }_{\mu\nu}(X) \; T^{ }_{\alpha\beta}(0) 
   \bigr\rangle
 \;, \la{covariant}
\ee
where $D$ denotes the dimension of space-time, 
$
 X \; \equiv \; (\tau,\vec{x})
$, 
and 
$
 \tau \in ( 0 , \frac{1}{T} )
$.
Here we have defined the projector 
($
 L^{ }_{\mu\nu;\alpha\beta}
 L^{ }_{\alpha\beta;\gamma\delta}
 = 
 L^{ }_{\mu\nu;\gamma\delta}
$)
\be
 L^{ }_{\mu\nu;\alpha\beta}
 \; \equiv \;  
 \frac{
   \PT_{\mu\alpha} 
   \PT_{\nu\beta} 
 + 
   \PT_{\mu\beta} 
   \PT_{\nu\alpha} 
 }{2} 
 - 
 \frac{ 
   \PT_{\mu\nu} 
   \PT_{\alpha\beta} 
 }{D-2}
 \;, \quad
 \PT_{\mu\nu} 
 \; \equiv \; 
 \delta^{ }_{\mu i}\delta^{ }_{\nu j}
 \biggl(
  \delta^{ }_{ij} - \frac{k^{ }_i k^{ }_j}{k^2} 
 \biggr)
 \;, \la{defL}
\ee
which is symmetric
($
  L^{ }_{\mu\nu;\alpha\beta} 
= L^{ }_{\nu\mu;\alpha\beta}
= L^{ }_{\alpha\beta;\mu\nu}
$)
and projects
onto transverse
($
  K^{ }_\mu L^{ }_{\mu\nu;\alpha\beta} = 
  k^{ }_i \delta^{ }_{i\mu} L^{ }_{\mu\nu;\alpha\beta} = 0 
$)
and traceless 
($
  \delta^{ }_{\mu\nu}L^{ }_{\mu\nu;\alpha\beta} = 0
$)
modes.
We also denote
\be
 \PT_{\vec{p}} 
 \; \equiv \; 
 \PT_{\mu\nu} P^{ }_\mu P^{ }_\nu 
 \; = \; 
 p^2 - \frac{(\vec{p}\cdot\vec{k})^2}{k^2}
 \;. 
\ee

As $T^{ }_{\mu\nu}$ we take the Standard Model energy-momentum tensor, 
which we write in Euclidean metric. 
Given that $L^{ }_{\mu\nu;\alpha\beta}$ projects out trace parts, it is 
enough to include non-trace ones, 
\ba
 T^{ }_{\mu\nu} & \supset & 
 F^{a_i}_{\mu\alpha} F^{a_i}_{\nu\alpha}
 + 
 (D^{ }_\mu\phi)^\dagger (D^{ }_\nu \phi) 
 + 
 (D^{ }_\nu\phi)^\dagger (D^{ }_\mu \phi) 
 \nn 
 & + & 
 \frac{1}{4} \Bigl[ 
 \bar{q}^{ }_\rmii{L} 
 \bigl( 
   \gamma^{ }_\mu\! \overleftrightarrow{D}_{\!\nu}^{ }
 + 
   \gamma^{ }_\nu\! \overleftrightarrow{D}_{\!\mu}^{ }
 \bigr) q^{ }_\rmii{L} 
 + 
 \bar{u}^{ }_\rmii{R} 
 \bigl( 
   \gamma^{ }_\mu\! \overleftrightarrow{D}_{\!\nu}^{ }
 + 
   \gamma^{ }_\nu\! \overleftrightarrow{D}_{\!\mu}^{ }
 \bigr) u^{ }_\rmii{R} 
 + 
 \bar{d}^{ }_\rmii{R} 
 \bigl( 
   \gamma^{ }_\mu\! \overleftrightarrow{D}_{\!\nu}^{ }
 + 
   \gamma^{ }_\nu\! \overleftrightarrow{D}_{\!\mu}^{ }
 \bigr) d^{ }_\rmii{R}
 \nn 
 & & \; + \, 
 \bar{\ell}^{ }_\rmii{L} 
 \bigl( 
   \gamma^{ }_\mu\! \overleftrightarrow{D}_{\!\nu}^{ }
 + 
   \gamma^{ }_\nu\! \overleftrightarrow{D}_{\!\mu}^{ }
 \bigr) \ell^{ }_\rmii{L} 
 + 
 \bar{\nu}^{ }_\rmii{R} 
 \bigl( 
   \gamma^{ }_\mu\! \overleftrightarrow{D}_{\!\nu}^{ }
 + 
   \gamma^{ }_\nu\! \overleftrightarrow{D}_{\!\mu}^{ }
 \bigr) \nu^{ }_\rmii{R} 
 + 
 \bar{e}^{ }_\rmii{R} 
 \bigl( 
   \gamma^{ }_\mu\! \overleftrightarrow{D}_{\!\nu}^{ }
 + 
   \gamma^{ }_\nu\! \overleftrightarrow{D}_{\!\mu}^{ }
 \bigr) e^{ }_\rmii{R}
 \Bigr] 
 \;, \la{Tmunu} \hspace*{5mm}
\ea
where the $a_i$ label the generators of the various gauge groups; 
$\phi$ is the Higgs doublet; 
$q^{ }_\rmii{L},\ell^{ }_\rmii{\,L}$ 
are the left-handed quark and lepton doublets, respectively; 
and 
$u^{ }_\rmii{R},d^{ }_\rmii{R},\nu^{ }_\rmii{R},e^{ }_\rmii{R}$
are the corresponding right-handed components. 
The covariant derivative has the form
\be
 D^{ }_\mu = 
 \partial^{ }_\mu - i g^{ }_1 Y A^{ }_\mu
 - i g^{ }_2 T^{a_2} A^{a_2}_\mu \aL 
 - i g^{ }_3 T^{a_3} A^{a_3}_\mu
 \;, \la{Dmu}
\ee 
where $g^{ }_1, g^{ }_2, g^{ }_3$ are gauge couplings, 
$\aL$ is the left-handed projector and the hypercharge 
assignments are 
$
 Y = -\frac{1}{2}$, 
$-\frac{1}{2\Nc}$, 
$- \frac{\Nc + 1}{2\Nc}$, 
$\frac{\Nc - 1}{2\Nc}$, 
$\frac{1}{2}$, 
$0$, 
$1$
for
$
 \phi, q^{ }_\rmii{L},u^{ }_\rmii{R},d^{ }_\rmii{R}, \ell^{ }_\rmii{\,L}, 
 \nu^{ }_\rmii{R}, e^{ }_\rmii{R}
$, 
respectively~\cite{nc1}. 
We note that because of their vanishing gauge charge assignments
and the omission of their Yukawa couplings, the fields $ \nu^{ }_\rmii{R} $
do not contribute to $2\leftrightarrow 2$ scatterings 
and have thus no effect on our final results
(traditionally, $\nu^{ }_\rmii{R}$ are often omitted from 
the outset). 

In order to avoid inverse polynomials of $D$
in \se\ref{ss:imag}, the result for 
$G^\rmii{E}_{12;12}$ is expressed as 
\ba
  && \hspace*{-1.5cm} G^\rmii{E}_{12;12}  \; \equiv \;  
  \frac{2}{D(D-2)(D-3)}
  \biggl\{ 
  \nn 
  & +  & 
  \nS^{ } \Phi^{ }_s 
  + 2 \nG^{ }(1 + \Nc^{ }) \Phi^{ }_f 
  + (2 + \Nc^{ }\CF^{ }) \Phi^{ }_g
  + \nS^{ }\lambda \, \Phi^{ }_{s(s)}
  + \bigl( 3 g_2^2 + \Nc^2 \CF^{ } g_3^2 \bigr) \, \Phi^{ }_{g(g)}
  \nn 
  & + &  \nS^{ } |h^{ }_t|^2 \Nc^{ }
    \Bigl[
       \Phi^{ }_{s(f)} 
     + \Phi^{ }_{f(s)} 
     + \Phi^{ }_{s|f} 
    \Bigr]
  + 
  \nS^{ }(g_1^2 + 3 g_2^2)
    \Bigl[
       \Phi^{ }_{s(g)} 
     + \Phi^{ }_{g(s)} 
     + \Phi^{ }_{s|g} 
    \Bigr]
  \nn 
  & + &
  \nG^{ }\,
  \biggl[ 
   \frac{(\Nc^{ }+1)(\Nc^{ }+2) g_1^2}{4\Nc^{ }} 
   + \frac{3 (\Nc^{ }+1) g_2^2}{4}
   + 2 \Nc^{ }\CF^{ } g_3^2
  \biggr] 
    \Bigl[
       \Phi^{ }_{f(g)} 
     + \Phi^{ }_{g(f)} 
     + \Phi^{ }_{f|g} 
    \Bigr]
  \nn 
  & + &
  \rmO(g^4)
  \biggr\} 
  \;, \hspace*{4mm} \la{Phi}
\ea
where $\nS^{ } = 1$ is the number of Higgs doublets, 
$\nG^{ }\equiv 3$ is the number of fermion generations, 
$\CF^{ }\equiv (\Nc^2 - 1)/(2\Nc^{ })$, 
and $\rmO(g^4)$ refers generically to any 3-loop contribution.\footnote{%
 The Higgs self-coupling and top Yukawa coupling appear in a Euclidean
 Lagrangian as $L^{ }_\E \supset \lambda(\phi^\dagger\phi)^2 
 + \bar{q}^{ }_\rmii{L} h^{ }_t t^{ }_\rmii{R}\tilde\phi
 + \tilde{\phi}^\dagger \bar{t}^{ }_\rmii{R} h_t^* q^{ }_\rmii{L}$, 
 whereas other Yukawa couplings are omitted. 
 } 
Here $s,f,g$ refer to effects from scalars, fermions, and gauge bosons, 
respectively; 
$\Phi^{ }_{a}$ is a 1-loop diagram with a particle of type $a$; 
$\Phi^{ }_{a(b)}$ is a 2-loop diagram where 
a particle of type $a$ couples to 
$T^{ }_{\mu\nu}$ and a particle of type $b$ appears in a loop; 
and $\Phi^{ }_{a|b}$ is a 2-loop diagram involving 
a cross correlation between the 
energy-momentum tensors of particles of types $a$ and $b$
(in terms of matrix elements this corresponds to an interference term). 
The corresponding Feynman diagrams are shown in 
\fig\ref{fig:graphs}. 

%
\begin{figure}[t]
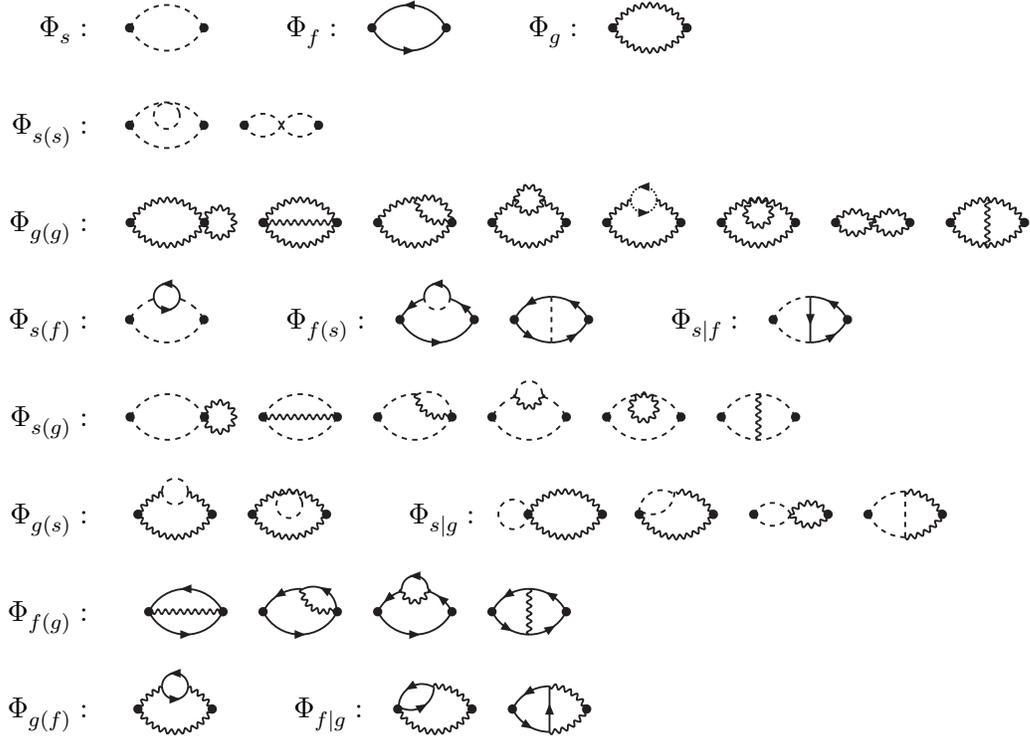


\begin{eqnarray*}
 \Phi^{ }_s: 
&& 
 \hspace*{-5mm}
 \Phis \;
 \Phi^{ }_f: \Phif \; 
 \Phi^{ }_g: \Phig \;
 \\[1mm]
 \Phi^{ }_{s(s)}: 
&& 
 \hspace*{-5mm}
 \PhissF \hspace*{-0.8cm}
 \PhissG 
 \\[1mm]
 \Phi^{ }_{g(g)}: 
&& 
 \hspace*{-2.5mm}
 \PhiggA \hspace*{-0.8cm} 
 \PhiggB \hspace*{-0.8cm}
 \PhiggC \hspace*{-0.8cm}
 \PhiggD \hspace*{-0.8cm}
 \PhiggE \hspace*{-0.8cm}
 \PhiggF \hspace*{-0.8cm}
 \PhiggG \hspace*{-0.8cm}
 \PhiggH  
 \\[1mm]
 \Phi^{ }_{s(f)}: 
&& 
 \hspace*{-5mm}
 \Phisf \;
 \Phi^{ }_{f(s)}: 
 \PhifsD \hspace*{-0.8cm}
 \PhifsH \; 
 \Phi^{ }_{s|f}: \Phisxf \;
 \\[1mm]
 \Phi^{ }_{s(g)}: 
&& 
 \hspace*{-2.5mm}
 \PhisgA  \hspace*{-0.8cm}
 \PhisgB  \hspace*{-0.8cm}
 \PhisgC  \hspace*{-0.8cm}
 \PhisgD  \hspace*{-0.8cm}
 \PhisgF  \hspace*{-0.8cm}
 \PhisgH  
 \\[1mm]
 \Phi^{ }_{g(s)}: 
&& 
 \hspace*{-5mm}
 \;
 \PhigsD \hspace*{-0.8cm} 
 \PhigsF \; 
 \Phi^{ }_{s|g}: 
 \PhisxgA  \hspace*{-0.6cm}
 \PhisxgC  \hspace*{-0.8cm}
 \PhisxgG  \hspace*{-0.8cm}
 \PhisxgH  
 \\[1mm]
 \Phi^{ }_{f(g)}: 
&& 
 \hspace*{-2.5mm}
 \PhifgB  \hspace*{-0.8cm}
 \PhifgC  \hspace*{-0.8cm}
 \PhifgD  \hspace*{-0.8cm}
 \PhifgH  
 \\[1mm]
 \Phi^{ }_{g(f)}: 
&& 
 \hspace*{-5mm}
 \;
 \PhigfD \; 
 \Phi^{ }_{f|g}: 
 \PhifxgC  \hspace*{-0.8cm}
 \PhifxgH  
\end{eqnarray*}

\caption[a]{\small 
The 1 and 2-loop graphs contributing to \eq\nr{Phi}. 
Each subset is gauge independent. 
Dashed lines denote scalars; 
solid lines fermions; 
wiggly lines gauge fields; 
dotted lines ghosts; 
blobs the operator $T^{ }_{\mu\nu}$. 
Graphs obtained by symmetrizations
have been omitted.} 
\la{fig:graphs}
\end{figure}
%

%
\subsection{Retarded energy-momentum correlator}
\la{ss:imag}

As the gravitational wave production rate is dominated 
by very high temperatures, we treat all particles as massless 
for the moment (the role of thermal masses is discussed 
in \se\ref{ss:htl} and 
in appendices~\ref{sss:htl_fermion} and \ref{ss:1to2}). 
Then the results for the correlators
can be expressed in terms of the ``master'' sum-integrals~\cite{asz}
\ba 
 J^{c}_{ab} \!\! & \equiv & \!\!
 \Tint{P} j^{c}_{ab}
 \;, \quad
 \tilde{J}^{c}_{ab} \; \equiv \;
 \Tint{\{P\}} \hspace*{-2mm} j^{c}_{ab}
 \;, \quad
 j^{c}_{ab} \; \equiv \; 
 \frac{[\PT_{ \vec{p} }]^c_{ }[K^2]^x_{ }} 
       {[P^2]^a_{ }
        [(K-P)^2]^b_{ }
       }
 \;, \la{J_def} \\[2mm] 
 I^{f\!gh}_{abcde} \!\! & \equiv & \!\!  
 \Tint{PQ} \hspace*{-2mm} {i}^{f\!gh}_{abcde}
 \;, \quad
 \tilde{I}^{f\!gh}_{abcde} \; \equiv \; 
 \Tint{P\{Q\}} \hspace*{-5mm} {i}^{f\!gh}_{abcde}
 \;, \quad
 \hat{I}^{f\!gh}_{abcde} \; \equiv \; 
 \Tint{\{P\}Q} \hspace*{-5mm} {i}^{f\!gh}_{abcde}
 \;, \quad
 \bar{I}^{f\!gh}_{abcde} \; \equiv \; 
 \Tint{\{PQ\}} \hspace*{-5mm} {i}^{f\!gh}_{abcde}
 \;, \nn[2mm] 
 {i}^{f\!gh}_{abcde} & \equiv & 
 \frac{[\PT_{ \vec{p} }]^f_{ }
       [\PT_{ \vec{q} }]^g_{ }
       [\PT_{ \vec{q-p} }]^h_{ }[K^2]^y_{ }} 
       {[P^2]^a_{ }
        [Q^2]^b_{ }
        [(Q-P)^2]^c_{ }
        [(K-P)^2]^d_{ }
        [(K-Q)^2]^e_{ }
       }
 \;, \la{I_def}
\ea
where $\{P\}$ denotes a fermionic Matsubara four-momentum. 
The indices 
$x \equiv a+b - c$ and 
$y \equiv a+b+c+d+e-f-g-h-2$ guarantee
the overall dimensionality GeV$^4$.
In the fermionic cases the representation is not unique; 
for the class of masters discussed in \se\ref{ss:2to2}, which have
a cut corresponding to a $2\leftrightarrow 2$ scattering, we have
ordered the indices such that $a,c,e$ are non-negative. 

The reduction of the energy-momentum tensor correlator to the
basis of \eqs\nr{J_def} and \nr{I_def} has been carried out with
a self-designed algorithm implemented in \textsc{FORM}~\cite{form}. 
After the use of
symmetries related to substitutions of integration variables, and
noting that terms with odd numbers 
of $\gamma^{ }_5$-matrices do not contribute at this order, 
the results read
\ba
 \Phi^{ }_s  & = & 4(D-3) J^{2}_{11} 
 \;, \la{Phi_s} \\[2mm]
 \Phi^{ }_f  & = & -4(D-3) \tilde{J}^{2}_{11} + 
 \frac{D(D-3)}{2}
 \, \bigl( 2 \tilde{J}^{1}_{10} - \tilde{J}^1_{11} \bigr) 
 \;, \la{Phi_f} \\[2mm]
 \Phi^{ }_g  & = &
 2(D-3) \biggl[ (D-2) J^{2}_{11} 
 + D \, \bigl( J^{1}_{11} - J^{1}_{10} \bigr)
 + \frac{D(D-2)}{8}
   \, \bigl( J^{0}_{11} - 2 J^{0}_{10} + 4 J^{0}_{00} \bigr)  
 \biggr] 
 \;, \hspace*{5mm}  \la{Phi_g} \\[2mm] 
 \Phi^{ }_{s(s)} & = & -48(D-3) I^{200}_{21010}
 \;, \la{Phi_ss} \\[2mm] 
 \Phi^{ }_{g(g)} & = & 
 \frac{D(D-2)(D-3)}{2} \bigl[
                         - I^{000}_{11111} 
                         - I^{010}_{12101}
                         - I^{100}_{21100} 
                         +3 I^{100}_{10101}  
                          -12 I^{010}_{12001}
 \nn & + & 
                          2 \bigl( 
                                   I^{010}_{121-21}
                                 - I^{000}_{11100}
                                 - I^{010}_{11001}
                                 - I^{000}_{11000} 
                                 - I^{100}_{20010} 
                            \bigr)
 \nn & + & 
                          4 \bigl( 
                                   I^{100}_{121-11}
                                  +I^{100}_{111-11}
                                  +I^{100}_{11101} 
                                  +I^{000}_{11101} 
                                  -I^{010}_{21100}
                                  -I^{100}_{11001}
                                  -I^{100}_{12001}
                            \bigr)
                       \bigr] 
 \nn & + &   
 2D(D-3) \bigl[
              4 I^{010}_{11101}
            - 2 I^{100}_{11111} 
            -   I^{001}_{11111}  
         \bigr]
  +    
 2D(D-6) \bigl[
              2 I^{101}_{11111}
            +   I^{110}_{11111}  
         \bigr]
 \nn & - &   
  (3D^2-16D+12) \bigl[
              2 I^{200}_{11111}
            +   I^{002}_{11111}  
                \bigr]
 - \frac{D(D-3)(3D-10)}{2} I^{100}_{11100}
 \nn & + & 
 D(D-2) \bigl[
              4 \bigl( 
                   I^{110}_{12101}
                  -I^{110}_{21100}
                \bigr)
            + 2 \bigl(
                   I^{101}_{12101}
                  -I^{200}_{12101}  
                  +I^{020}_{21100}
                  -I^{011}_{21100}
                \bigr) 
 \nn & + & 
              I^{200}_{21100}
            - I^{020}_{12101} 
        \bigr]
 +
 \frac{(D-2)^2(D-3)}{2} \bigl[
                            D \bigl( 
                              I^{000}_{12000}
                           -  I^{000}_{12001}
                              \bigr) 
                           -8 I^{020}_{12001} 
                        \bigr]
 \;, \\[2mm] 
 \Phi^{ }_{s(f)} & = &
 8(D-3) \bigl[ 
 2 \hat{I}^{020}_{12001} - \hat{I}^{020}_{11101}
 \bigr]
 \;, \\[2mm] 
 \Phi^{ }_{f(s)} & = &
 \frac{D(D-3)}{2}
          \bigl[  
         4\bigl(
              \bar{I}_{111-11}^{010}
             +\bar{I}_{10101}^{100}
             -\bar{I}_{10110}^{100}
             -\bar{I}_{11100}^{100}
             -\bar{I}_{20100}^{100}
             +\bar{I}_{20110}^{100}
             +\bar{I}_{21000}^{100}
             -\bar{I}_{21010}^{100} 
           \bigr)
 \nn & - & 
          2\bigl(
              \bar{I}_{10101}^{001}
             +\bar{I}_{11011}^{100}
           \bigr)
         +    \bar{I}_{11011}^{001} 
           \bigr]
 +16 (D-3) \bigl[ 
              \bar{I}_{20110}^{200}
             -\bar{I}_{21010}^{200}
           \bigr]
 -\frac{3D-8}{2}\, \bar{I}_{11111}^{002} 
 \nn 
 & + & 2(D-2) \bigl[
         4 \bigl(
              \bar{I}_{11011}^{101}
             -\bar{I}_{11101}^{101}
             -\bar{I}_{11101}^{110} 
           \bigr)
        +2\bigl(
              \bar{I}_{11101}^{002}
             +\bar{I}_{11101}^{020}
             +\bar{I}_{11101}^{200} 
             -\bar{I}_{11011}^{200}
           \bigr)
  \nn & + &
              \bar{I}_{11111}^{101}
             -\bar{I}_{11011}^{002}
           \bigr]
  +   (D-4) \bigl[      
           8  \bar{I}_{11101}^{011}
         - 4  \bar{I}_{11011}^{110}
             +\bar{I}_{11111}^{200}
             -\bar{I}_{11111}^{110}
           \bigr]
 \;, \\[2mm] 
 \Phi^{ }_{s|f} & = &
  2 (D-2) 
 \bigl[
    4 \bigl( \tilde{I}^{011}_{11101}
           + \tilde{I}^{101}_{11101}
      \bigr)
 - 2 \bigl(  \tilde{I}^{011}_{11111}
           + \tilde{I}^{200}_{11101}
           + \tilde{I}^{020}_{11101}
           + \tilde{I}^{002}_{11101}
     \bigr)
  +  \tilde{I}^{020}_{11111}
 \nn & + & 
     \tilde{I}^{002}_{11111} 
 \bigr]
 + 8(D-3) \hat{I}^{020}_{11101}
 - 4 \bigl[ \tilde{I}^{110}_{11111} 
          + \tilde{I}^{101}_{11111} \bigr]
 - 2(D-4) \bigl[ \tilde{I}^{200}_{11111} 
             + 4 \tilde{I}^{110}_{11101}\bigr]
 \;, \\[2mm]
 \Phi^{ }_{s(g)} & = & 
   \frac{D-2}{2}\bigl[ 
                       4 I^{101}_{11111}
                     - 2 I^{200}_{11111}
                     -   I^{002}_{11111}
                \bigr] 
 - (D-4) I^{110}_{11111} 
 \nn 
 & + &
  \frac{D-3}{2}\bigl[
                  3 D I^{010}_{10101} 
                  - 8 I^{020}_{11101}
               -4(D-1)I^{200}_{21010}
                \bigr]
 \;, \\[2mm]
 \Phi^{ }_{g(s)} & = & 
  \frac{D(D-3)}{4}
  \bigl[ 
     4 \bigl(
       I^{100}_{121-11}
    -  I^{100}_{111-11}
    +  I^{010}_{11010}
    -  I^{010}_{21100}
    -  I^{010}_{21010} \bigr) 
 \nn & + &
     2 \bigl(
       I^{010}_{121-21}
    +  I^{010}_{11101}
    -  I^{100}_{20010} \bigr)  
    -  I^{100}_{21100}
    -  I^{010}_{12101}
    -  I^{010}_{10101}
    +6 I^{100}_{11010} 
 \nn & + &
     7 I^{100}_{11100}
   -12 I^{100}_{21010} 
  \bigr]
 + 2(D-2) \bigl[
            I^{200}_{11101}
          - I^{101}_{11101} 
         \bigr]
 - 4 I^{110}_{11101} 
 - (D-4) I^{020}_{11101}  
\nn 
& + &
 \frac{D}{2} \bigl[
               4 \bigl( 
                    I^{110}_{12101} 
                   -I^{110}_{21100}
                 \bigr)
               + 2 \bigl(
                    I^{101}_{12101} - I^{200}_{12101} 
                   +I^{020}_{21100} - I^{011}_{21100}
                   \bigr)
               + I^{200}_{21100}
               - I^{020}_{12101}               
             \bigr]
\nn 
& + &
  \frac{(D-2)(D-3)}{4}
        \bigl[
             D \bigl( 2 I^{000}_{11010} 
                       +I^{000}_{21000}  
                       -I^{000}_{21010}
                       -4 I^{000}_{11000}
                \bigr)
            -8  I^{200}_{21010} 
        \bigr]
 \;, \\[2mm]
 \Phi^{ }_{s|g} & = & 
  (D-2) \bigl[
          2 \bigl(
                  I^{110}_{11111}
                + I^{101}_{11101}
        - I^{200}_{11111}     
                - I^{200}_{11101} 
            \bigr)
        -  I^{002}_{11111}  
                 \bigr]
 + 4 I^{101}_{11111}
\nn &+ & 
  \frac{D(D-3)}{2} \bigl[ 
           4 \bigl( 
                 I^{100}_{111-11}
               - I^{100}_{11010}
               - I^{010}_{11010} 
             \bigr) 
         +2 I^{100}_{11101}
          -  I^{010}_{11101}
          -  I^{010}_{10101} 
          - 5 I^{100}_{11100}
                   \bigr]
\nn &+ &
          4 I^{110}_{11101}
         + (5D-16) I^{020}_{11101}
  +\frac{D(D-2)(D-3)}{2}
        \bigl[
          2 I^{000}_{11000}
          - I^{000}_{11010}  
        \bigr]  
 \;, \\[2mm]
 \Phi^{ }_{f(g)} & = & 
 \frac{D(D-2)(D-3)}{2} \bigl[ 
                         2  \bigl( 
                              \bar{I}^{010}_{111-11}
                             +\bar{I}^{000}_{101-11}
                             +\bar{I}^{010}_{02101}
                             -\bar{I}^{010}_{12001}
                             -\bar{I}^{010}_{01101}
                             +\bar{I}^{010}_{12000} 
                             -\bar{I}^{010}_{02100} 
                            \bigr)
 \nn 
 & - & 
                           \bar{I}^{000}_{10101} 
                         - \bar{I}^{000}_{01010} 
                       \bigr]
  +  
 \frac{D(D-3)}{2}\bigl[ 
                    \bar{I}^{100}_{11111} 
                  + \bar{I}^{010}_{11111} 
                  - \bar{I}^{001}_{11111} 
                 + 2\bar{I}^{001}_{11100} 
               - 2 D\bar{I}^{010}_{11100}
                 \bigr]
 \nn 
 & + & 
 \frac{(D-4)(D+2)}{4} \bigl[
                      \bar{I}^{200}_{11111} 
                    + \bar{I}^{020}_{11111}  
                      \bigr]
   -\frac{3D^2-18D+32}{4} \bar{I}^{002}_{11111} 
   -\frac{D^2-18D+40}{2} \bar{I}^{110}_{11111}
 \nn 
  & + & 
 (D-2)^2 \bigl[
           2 \bigl(
               \bar{I}^{200}_{11101}  
              +\bar{I}^{002}_{11101}  
              +\bar{I}^{101}_{11011} 
              +\bar{I}^{011}_{11011}  
             \bigr) 
  -  
               \bar{I}^{200}_{11011} 
             - \bar{I}^{020}_{11011} 
             - \bar{I}^{002}_{11011} 
           -4  \bar{I}^{101}_{11101}  
         \bigr] 
 \nn
 & + &
  \frac{D(D-3)(D-10)}{2} \bigl[
                           2 \bar{I}^{100}_{10101} 
                           - \bar{I}^{001}_{10101} 
                         \bigr]
 -\frac{D(D-3)(D-6)}{4} \bigl[ 
                           \bar{I}^{100}_{11011}
                         + \bar{I}^{010}_{11011} 
                         - \bar{I}^{001}_{11011} 
                        \bigr]
 \nn
 & + &
  2(D^2+4D-20)\bar{I}^{020}_{11101} 
 -2(D-2)(D-4) \bar{I}^{110}_{11011} 
 +     \frac{D^2-8D+20}{2} \bigl[
                           \bar{I}^{101}_{11111} 
                         + \bar{I}^{011}_{11111} 
                        \bigr]
 \nn 
 & + & 
  4(D^2-10D+20)\bar{I}^{011}_{11101}
 - 4 (D-4)^2  \bar{I}^{110}_{11101}
 +  8(D-2)(D-3)\bigl[
                 \bar{I}^{020}_{02101}
                -\bar{I}^{020}_{12001} 
              \bigr]
 \;, \nn \\[2mm]
 \Phi^{ }_{g(f)} & = & 
 \frac{D(D-2)(D-3)}{2} \bigl[ 
                           \tilde{I}^{000}_{11100} 
                         - \hat{I}^{000}_{11101} 
                         - \tilde{I}^{000}_{101-11}
 \nn & + & 
                         2 \bigl( 
                              \tilde{I}^{000}_{10101} 
                             +\tilde{I}^{000}_{21010}
                             -\tilde{I}^{000}_{21000} 
                           \bigr)
                         -3\tilde{I}^{000}_{01100}
                         -4\tilde{I}^{000}_{11010}
                         +8\tilde{I}^{000}_{11000} 
                       \bigr]
\nn
 & + & 
  D(D-3) \bigl[ 
             2 \bigl(
                  \hat{I}^{100}_{111-11} 
                 -\hat{I}^{001}_{111-11}
                 +\hat{I}^{001}_{121-11}
                 -\hat{I}^{010}_{121-21}
                 -\hat{I}^{100}_{121-11}
                 -\hat{I}^{100}_{12101}
 \nn & - & 
                  \tilde{I}^{001}_{10101}
                 -\tilde{I}^{001}_{21100}
                 +\tilde{I}^{001}_{21000}
                 -\tilde{I}^{100}_{21000}
                 -\tilde{I}^{010}_{21000}
               \bigr)
                +  \hat{I}^{010}_{12101}
               +\tilde{I}^{100}_{21100}
            +3 \bigl( 
                  \tilde{I}^{100}_{10101} 
                 +\tilde{I}^{100}_{11100} 
                \bigr)
 \nn & + & 
             4 \bigl( 
                  \hat{I}^{100}_{11101}
               -\tilde{I}^{100}_{11010} 
               +\tilde{I}^{010}_{21100}
               \bigr)
            -6 \bigl( 
                  \hat{I}^{010}_{11101}
              + \tilde{I}^{010}_{11100} 
               \bigr)  
           +10 \tilde{I}^{100}_{21010} 
        \bigr]
\nn
 & + & 
  2D \bigl[ 
             4 \bigl( 
                    \tilde{I}^{110}_{21100}
                    - \hat{I}^{110}_{12101}
               \bigr)
           + 2 \bigl( 
                 \hat{I}^{200}_{12101}
                -\hat{I}^{101}_{12101} 
              +\tilde{I}^{011}_{21100} 
              -\tilde{I}^{020}_{21100} 
               \bigr)
           +  \hat{I}^{020}_{12101}
           -\tilde{I}^{200}_{21100} 
    \bigr]
\nn
 & + & 
  16 \hat{I}^{110}_{11101}
 +8(D-2) \bigl[
            \hat{I}^{101}_{11101}
          - \hat{I}^{200}_{11101}  
         \bigr]
 \nn & + & 
  8(D-2)(D-3) \tilde{I}^{200}_{21010}
 -4(D^2-6D+10) \hat{I}^{020}_{11101} 
 \;, \\[2mm]
 \Phi^{ }_{f|g} & = & 
 D(D-2)(D-3) \bigl[ 
              \tilde{I}^{000}_{101-11}
            - \tilde{I}^{000}_{10101} 
            + \tilde{I}^{000}_{01100}
            + 2 \tilde{I}^{000}_{11010}
            - 4 \tilde{I}^{000}_{11000} 
             \bigr]
 \nn & + & 
 D(D-3) \bigl[ 
            \tilde{I}^{010}_{11111} 
           +\tilde{I}^{001}_{11111} 
           -\tilde{I}^{100}_{11111} 
          -2\tilde{I}^{100}_{11100} 
          +3\tilde{I}^{010}_{10101} 
          +5\tilde{I}^{001}_{10101} 
          -6\tilde{I}^{100}_{10101} 
\nn & + & 
           4 \bigl( 
                \hat{I}^{001}_{111-11}
               -\hat{I}^{100}_{111-11} 
               +\hat{I}^{010}_{11101} 
             -\tilde{I}^{010}_{11101} 
             +\tilde{I}^{100}_{11010} 
              \bigr)
          + 8 \bigl( 
             \tilde{I}^{010}_{11100} 
            - \hat{I}^{100}_{11101}
              \bigr)
        \bigr]
 \nn & + & 
 2(D^2 - 10 D + 20)
         \bigl[
          2 \bigl( 
             \tilde{I}^{110}_{11011}
          -  \tilde{I}^{110}_{11101}
             \bigr)
          -\tilde{I}^{011}_{11111}   
         \bigr]
 + (D^2-2D-4) \bigl[ 
           \tilde{I}^{002}_{11111} 
          +\tilde{I}^{020}_{11111} 
            \bigr]
 \nn & + &
    8(D-2) \bigl[
             \hat{I}^{200}_{11101}
           - \hat{I}^{101}_{11101}  
           \bigr]
 + 4(D^2-6D+10) \hat{I}^{020}_{11101}
 -16 \hat{I}^{110}_{11101}
 \nn & + & 
   2(D-2)^2 \bigl[
            2 \bigl( 
                 \tilde{I}^{101}_{11101} 
               - \tilde{I}^{101}_{11011}
              \bigr)
            +\tilde{I}^{002}_{11011}
            -\tilde{I}^{002}_{11101}
            +\tilde{I}^{200}_{11011}
            -\tilde{I}^{200}_{11101}  
             \bigr]
 \nn & + &
 2(D^2-12D+28)\tilde{I}^{020}_{11011}  
 -2(D^2+4D-20)\tilde{I}^{020}_{11101}
 \nn & + &
   (D-4)^2 \bigl[
          4 \bigl( 
                \tilde{I}^{011}_{11101} 
              - \tilde{I}^{011}_{11011} 
            \bigr)
           -\tilde{I}^{200}_{11111} 
            \bigr]
   -2(3D-10)\bigl[
           \tilde{I}^{101}_{11111}
          +\tilde{I}^{110}_{11111}  
            \bigr] 
 \;. \la{Phi_fxg}
\ea
The computation was carried out in a general covariant gauge, and
we have checked that the gauge parameter drops out exactly. 
The result for $\Phi^{ }_{g(g)}$ can 
be crosschecked against ref.~\cite{asz}. 

%
\subsection{Extracting $2\leftrightarrow 2$ cuts at light cone}
\la{ss:2to2}

As discussed below \eq\nr{master}, from each $\Phi$ we need to extract
the cut
$
 \im \Phi |^{ }_{k^{ }_n\to -i [k + i 0^+]}
$.
For the moment we only consider the cuts corresponding to 
$2\leftrightarrow 2$ scatterings, which originate from 
the masters $I$, with the discussion of 
$1\leftrightarrow 2$ reactions postponed to appendix~\ref{ss:1to2}. 
As we restrict ourselves to the light cone, structures
which have a positive power $y$ in \eq\nr{I_def} yield no contribution.
This implies that the only structures playing a role are of the types
\be
 I^{000}_{101-11}\;, \quad
 I^{100}_{10101}\;, \quad
 I^{100}_{111-11}\;, \quad
 I^{100}_{121-21}\;, \quad
 I^{200}_{11101}\;. \la{cut_diags}
\ee 

We denote the phase space of $2\leftrightarrow 2$ scatterings by
\be
 \int \! {\rm d}\Omega^{ }_{2\to2} 
 \; \equiv \; 
 \int\! \frac{{\rm d}^3\vec{p}^{ }_1}{(2\pi)^3 2 p^{ }_1}
 \int\! \frac{{\rm d}^3\vec{p}^{ }_2}{(2\pi)^3 2 p^{ }_2}
 \int\! \frac{{\rm d}^3\vec{k}^{ }_1}{(2\pi)^3 2 k^{ }_1}
 \, 
 (2\pi)^4 \delta^{(4)}(
  \mathcal{P}^{ }_1 + \mathcal{P}^{ }_2 
 - \mathcal{K}^{ }_1 - \mathcal{K}^{ }_2)
 \;, 
\ee
where $\mathcal{P}^{ }_i \equiv (p^{ }_i,\vec{p}^{ }_i)$ 
with $p^{ }_i \equiv |\vec{p}^{ }_i|$, and 
$\mathcal{K}^{ }_2 \equiv \mathcal{K} \equiv (k,\vec{k})$.
Distribution functions are denoted by 
\be
 n^{ }_\sigma(\epsilon) \; \equiv \; \frac{\sigma}{e^{\epsilon/T} - \sigma}
 \;, \quad
 \sigma = \pm
 \;, \la{n_sigma}
\ee
so that $n^{ }_+ = \nB^{ }$ and $n^{ }_- = - \nF^{ }$ are the Bose
and Fermi distributions, respectively. Distribution functions appear
in the combination 
\ba
 \mathcal{N}^{ }_{\tau_1;\sigma_1\sigma_2}
 & \equiv & n^{ }_{\tau_1}(k^{ }_1) 
 \,[1 + n^{ }_{\sigma_1} (p^{ }_1)]  
 \,[1 + n^{ }_{\sigma_2} (p^{ }_2)]
 -    n^{ }_{\sigma_1} (p^{ }_1)
      \,n^{ }_{\sigma_2} (p^{ }_2) 
 \,[1 + n^{ }_{\tau_1}(k^{ }_1)]
 \;. \la{phasespace}
\ea
Mandelstam variables are defined as usual, 
$ 
 s \equiv (\mathcal{P}^{ }_1 + \mathcal{P}^{ }_2)^2
$, 
$
 t \equiv (\mathcal{P}^{ }_1 - \mathcal{K}^{ }_1)^2
$, 
$
 u \equiv (\mathcal{P}^{ }_2 - \mathcal{K}^{ }_1)^2
$.

With this notation, the $2\leftrightarrow 2$ cuts for the structures 
in \eq\nr{cut_diags} read
\ba
 \im \bigl\{ I^{f\!gh}_{1 b 1 d 1} \bigr\} 
 \bigr|^{2\leftrightarrow 2}_{k_n\to -i[k+i0^+]} 
 & = &  
 \frac{1}{2} 
 \int \! {\rm d}\Omega^{ }_{2\to2} \, 
 \biggl\{
  \frac{
  [\PT_{ \vec{k}^{ }_1 }]^f_{ }
  [\PT_{ \vec{p}^{ }_1 }]^g_{ }
  [\PT_{ \vec{p}^{ }_2 }]^h_{ }
  \mathcal{N}^{ }_{\sigma_a;\sigma_e\sigma_c} 
  }{[-u]^b\,[-s]^d} 
 \nn 
 &  & \hspace*{1.8cm} + \, 
  \frac{
  [\PT_{ \vec{p}^{ }_1 }]^f_{ }
  [\PT_{ \vec{p}^{ }_2 }]^g_{ }
  [\PT_{ \vec{k}^{ }_1 }]^h_{ }
  \mathcal{N}^{ }_{\sigma_c;\sigma_a\sigma_e}   
  }{[-t]^b\,[-u]^d} 
 \nn 
 \Cut
 &  & \hspace*{1.8cm} + \,
  \frac{
  [\PT_{ \vec{p}^{ }_2 }]^f_{ }
  [\PT_{ \vec{k}^{ }_1 }]^g_{ }
  [\PT_{ \vec{p}^{ }_1 }]^h_{ }
 \mathcal{N}^{ }_{\sigma_e;\sigma_c\sigma_a}   
  }{[-s]^b\,[-t]^d} 
 \biggr\}
 \;, \hspace*{7mm} \la{cut}
\ea
where $\sigma^{ }_a, \sigma^{ }_c$ and $\sigma^{ }_e$ 
label the statistics of the 1$^\rmi{st}_{ }$, 3$^\rmi{rd}_{ }$ 
and 5$^\rmi{th}_{ }$ subscript of $I$, respectively. 
The diagram illustrates the cuts, with crosses on the 
propagators $b$ and $d$ of which at least one 
comes with a zero or negative power.

We can now collect together the cuts from \eqs\nr{Phi_ss}--\nr{Phi_fxg}.
In so doing we also set $D\to 4$ for simplicity, as there are 
no ultraviolet divergences in these cuts. Denoting by $\mathcal{C}$
an operation which produces an integrand for \eq\nr{cut}, 
{\it viz.}
\be
 \lim_{D\to 4}
 \im \bigl\{ \Phi \bigr\} 
 \bigr|^{2\leftrightarrow 2}_{k_n\to -i[k+i0^+]} 
 \; \equiv \;  
 \int \! {\rm d}\Omega^{ }_{2\to2} \, 
 \mathcal{C} \Phi
 \;, \la{C_Phi}
\ee
and making use of symmetries such as 
$
 \tilde{I}^{f\!gh}_{1b101} = \bar{I}^{hg\!f}_{1b101}
$ 
(obtained by the substitution $P\to Q-P$), 
the non-zero contributions for the 
combinations appearing in \eq\nr{Phi} read
\ba
 \mathcal{C} \Phi^{ }_{g(g)} & = & 
 4 \mathcal{C} 
    \Bigl[
       \Phi^{ }_{s(g)} 
     + \Phi^{ }_{g(s)} 
     + \Phi^{ }_{s|g} 
    \Bigr]
 \la{cut_gg} \\[2mm] 
 & = &   
 4 \mathcal{C} \bigl[ 
                        2 I^{010}_{121-21} + 4 I^{100}_{111-11}
                      + 3 I^{100}_{10101}
               \bigr]
 \nn & = & 
 2 \, \mathcal{N}^{ }_{+;++} \,
 \biggl\{ 
   \PT_{ \vec{p}^{ }_1 } \,
   \biggl( 3 + \frac{4u}{t} + \frac{2s^2}{u^2}\biggr) 
 + 
   \PT_{ \vec{p}^{ }_2 } \,
   \biggl( 3 + \frac{4t}{s} + \frac{2u^2}{t^2}\biggr) 
 \nn & + &  
   \PT_{ \vec{k}^{ }_1 } \,
   \biggl( 3 + \frac{4s}{u} + \frac{2t^2}{s^2}\biggr) 
 \biggr\}
 \;, \la{cut_sxg} \\[2mm] 
 \mathcal{C}
    \Bigl[
       \Phi^{ }_{s(f)} 
     + \Phi^{ }_{f(s)} 
     + \Phi^{ }_{s|f} 
    \Bigr]
  & = & 
 4 \mathcal{C} 
  \bigl[ 
    2 \bigl( 
        \bar{I}^{010}_{111-11}
       +\bar{I}^{100}_{10101}
      \bigr) 
       -\bar{I}^{001}_{10101} 
  \bigr]
 \nn & = & 
 2  \, \mathcal{N}^{ }_{-;-+} \, \biggl\{ 
    \frac{2 s\, \PT_{ \vec{p}^{ }_1 }}{u}
  +  2 \PT_{ \vec{k}^{ }_1 }
  -  \PT_{ \vec{p}^{ }_2 }   
 \biggr\}
 \nn & + & 
 2  \, \mathcal{N}^{ }_{-;+-} \, \biggl\{ 
    \frac{2 t\, \PT_{ \vec{k}^{ }_1 }}{s}
  +  2 \PT_{ \vec{p}^{ }_2 } 
  -  \PT_{ \vec{p}^{ }_1 }
 \biggr\}
 \nn & + & 
 2  \, \mathcal{N}^{ }_{+;--} \, \biggl\{ 
    \frac{2 u\, \PT_{ \vec{p}^{ }_2 }}{t}
  + 2 \PT_{ \vec{p}^{ }_1 }
  - \PT_{ \vec{k}^{ }_1 } 
 \biggr\}
 \;, \la{cut_sxf} \\[2mm] 
 \mathcal{C}
    \Bigl[
       \Phi^{ }_{f(g)} 
     + \Phi^{ }_{g(f)} 
     + \Phi^{ }_{f|g} 
    \Bigr]
  & = &
 4\mathcal{C}\bigl[ 
     2\bigr( 
         \hat{I}^{001}_{111-11}
        -\hat{I}^{100}_{111-11} 
        -\hat{I}^{010}_{121-21} 
        +\bar{I}^{010}_{111-11}
        +\bar{I}^{000}_{101-11} 
      \bigl) 
  + \tilde{I}^{000}_{101-11}
 \bigr]
 \nn & = &  
 4  \, \mathcal{N}^{ }_{-;-+} \, \biggl\{ 
    \frac{s\, \PT_{ \vec{p}^{ }_1 }}{u}
  + 
    \frac{u\, \bigl[ \PT_{ \vec{k}^{ }_1 } - \PT_{ \vec{p}^{ }_1 } \bigr] }{t}
  - \frac{u^2\, \PT_{ \vec{p}^{ }_2 } }{t^2}
 \biggr\} 
 \nn & + &  
 4  \, \mathcal{N}^{ }_{-;+-} \, \biggl\{ 
    \frac{t\, \PT_{ \vec{k}^{ }_1 }}{s}
  + 
    \frac{s\, \bigl[ \PT_{ \vec{p}^{ }_2 } - \PT_{ \vec{k}^{ }_1 } \bigr] }{u}
  - \frac{s^2\, \PT_{ \vec{p}^{ }_1 } }{u^2}
 \biggr\} 
 \nn & + &  
 4  \, \mathcal{N}^{ }_{+;--} \, \biggl\{ 
    \frac{u\, \PT_{ \vec{p}^{ }_2 }}{t}
  + 
    \frac{t\, \bigl[ \PT_{ \vec{p}^{ }_1 } - \PT_{ \vec{p}^{ }_2 } \bigr] }{s}
  - \frac{t^2\, \PT_{ \vec{k}^{ }_1 } }{s^2}
 \biggr\} 
 \;. \la{cut_fxg}
\ea

At the light cone, there is a further identity that has 
not been employed yet and that permits for a remarkable simplification
of \eqs\nr{cut_sxg}--\nr{cut_fxg}. Noting that for massless particles
$
 u = 2(\vec{k}\cdot\vec{p}^{ }_1 - k p^{ }_1)
$, 
and recalling that 
$
 \PT_{ \vec{p}^{ }_1 } =
 (k p^{ }_1 -  \vec{k}\cdot\vec{p}^{ }_1 )
 (k p^{ }_1 +  \vec{k}\cdot\vec{p}^{ }_1 ) / k^2 
$, 
we can make use of energy-momentum conservation to verify that 
\be
 \frac{\PT_{ \vec{k}^{ }_1 }}{s}
 + \frac{\PT_{ \vec{p}^{ }_2 }}{t}
 + \frac{\PT_{ \vec{p}^{ }_1 }}{u}
 = 
 -1
 \;. \la{magic} 
\ee
With this identity, combined with renamings 
$\vec{p}^{ }_1 \leftrightarrow \vec{p}^{ }_2$ as well as 
a repeated use of $s + t + u = 0$, all projectors $\PT_{ }$
can be eliminated, and the cuts
in \eqs\nr{cut_gg}--\nr{cut_fxg} can be written in a form
where the breaking of Lorentz invariance through the medium
manifests itself only through the distribution
functions $\mathcal{N}^{ }_{\tau_1;\sigma_1\sigma_2}$: 
\ba
 && \hspace*{-5cm}
 \mathcal{C} \Phi^{ }_{g(g)} \; = \;  
 4 \mathcal{C} 
    \Bigl[
       \Phi^{ }_{s(g)} 
     + \Phi^{ }_{g(s)} 
     + \Phi^{ }_{s|g} 
    \Bigr]
 \; = \;    
 2 \, \mathcal{N}^{ }_{+;++} \,
 \biggl\{ -2 \biggl( 
      \frac{s^2 + u^2}{t}
   +  \frac{t^2}{s} \biggr)
 \biggr\}
 \;, \hspace*{6mm} \la{cut_gg_2} \\[2mm] 
 \mathcal{C}
    \Bigl[
       \Phi^{ }_{s(f)} 
     + \Phi^{ }_{f(s)} 
     + \Phi^{ }_{s|f} 
    \Bigr]
 \!\! & = & \!\! 
 2  \, \mathcal{N}^{ }_{-;-+} \, \bigl\{ 2 t 
 \bigr\}
 +  
 2  \, \mathcal{N}^{ }_{+;--} \, \bigl\{ s 
 \bigr\}
 \;, \la{cut_sxf_2} \\[2mm] 
 \mathcal{C}
    \Bigl[
       \Phi^{ }_{f(g)} 
     + \Phi^{ }_{g(f)} 
     + \Phi^{ }_{f|g} 
    \Bigr]
 \!\! & = & \!\!
 4  \, \mathcal{N}^{ }_{-;-+} \, \biggl\{ 
    \frac{s^2 + u^2}{t}
 \biggr\} 
 +   
 4  \, \mathcal{N}^{ }_{+;--} \, \biggl\{ 
    \frac{t^2}{s}
 \biggr\} 
 \;. \la{cut_fxg_2}
\ea
We note that \eq\nr{cut_gg_2} could be written in a more symmetric form, 
but for later convenience we prefer to use the same structures 
as in \eq\nr{cut_fxg_2}.
Eqs.~\nr{cut_gg_2}--\nr{cut_fxg_2}
correspond to amplitudes squared for processes illustrated
in \fig\ref{fig:scat} (cf.\ \se\ref{ss:boltzmann}). 

%
\begin{figure}[t]
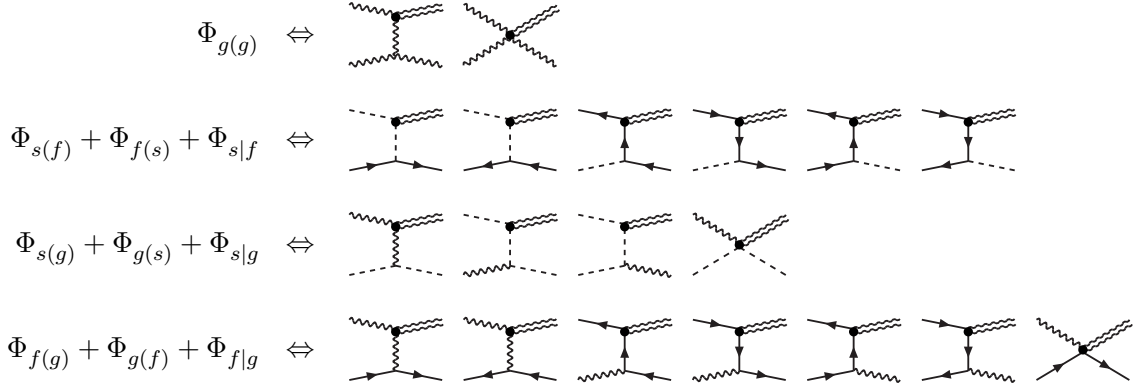


\vspace*{-1.0cm}%

\begin{eqnarray*}
 \Phi^{ }_{g(g)} 
 & \Leftrightarrow &
 \hspace*{0cm}
 \GraphggA
 \hspace*{-0.80cm}
 \GraphggB
 \nn[2mm]
       \Phi^{ }_{s(f)} 
     + \Phi^{ }_{f(s)} 
     + \Phi^{ }_{s|f} 
 & \Leftrightarrow &
 \hspace*{0cm}
 \GraphsfA
 \hspace*{-0.80cm}
 \GraphsfB
 \hspace*{-0.80cm}
 \GraphsfC
 \hspace*{-0.80cm}
 \GraphsfE
 \hspace*{-0.80cm}
 \GraphsfD
 \hspace*{-0.80cm}
 \GraphsfF
 \nn[2mm]
       \Phi^{ }_{s(g)} 
     + \Phi^{ }_{g(s)} 
     + \Phi^{ }_{s|g} 
 & \Leftrightarrow &
 \hspace*{0cm}
 \GraphsgA
 \hspace*{-0.80cm}
 \GraphsgB
 \hspace*{-0.80cm}
 \GraphsgC
 \hspace*{-0.80cm}
 \GraphsgD
 \nn[2mm]
       \Phi^{ }_{f(g)} 
     + \Phi^{ }_{g(f)} 
     + \Phi^{ }_{f|g} 
 & \Leftrightarrow &
 \hspace*{0cm}
 \GraphfgA
 \hspace*{-0.80cm}
 \GraphfgB
 \hspace*{-0.80cm}
 \GraphfgC
 \hspace*{-0.80cm}
 \GraphfgE
 \hspace*{-0.80cm}
 \GraphfgD
 \hspace*{-0.80cm}
 \GraphfgF
 \hspace*{-0.80cm}
 \GraphfgH
\end{eqnarray*}

\vspace*{-0.4cm}%

\caption[a]{\small 
 $t$-channel $2\leftrightarrow2$ scatterings contributing 
 to gravitational wave production (further processes are
 obtained with $u$ and $s$-channel reflections).
 The notation is as in \fig\ref{fig:graphs}, with the double 
 line indicating a graviton. 
 Up to numerical prefactors, 
 the amplitudes squared originating from 
 these processes, after summing over the physical polarization
 states of the gravitons and Standard Model
 particles, correspond to the cuts
 shown in \eqs\nr{cut_gg_2}--\nr{cut_fxg_2} (cf.\ \se\ref{ss:boltzmann}).
} 
\la{fig:scat}
\end{figure}
%

The drastic simplification that we have observed when going
on the light-cone has a known precedent: it also takes place for 
photon production from a thermal medium. Furthermore, in that case it is 
well understood. The transverse correlator to which physical 
photons couple, $\im G^\rmii{R}_\rmii{T}$, can be replaced by the
full vector correlator, 
$
 \im G^\rmii{R}_\rmii{V} = 
 \im G^\rmii{R}_\rmii{T} + \im G^\rmii{R}_\rmii{L}
$, because a Ward identity guarantees the
vanishing of $\im G^\rmii{R}_\rmii{L}$ for zero virtuality. We are 
not aware of a similar operator relation between the tensor channel
correlator in \eq\nr{covariant}
and one without any $\PT_{ }$'s, even if intriguing
relations between photon and graviton production 
amplitudes are known to exist (cf.\ \se\ref{ss:boltzmann}). 

%
\subsection{Connection to Boltzmann equations}
\la{ss:boltzmann}

The $2\leftrightarrow 2$ cuts of \se\ref{ss:2to2} 
can also be obtained from kinetic theory
and Boltzmann equations. As a starting point, we may, for $k\sim \pi T$,
write the leading-order contribution to eq.~\eqref{rate_gen} as
\begin{equation}
  \label{kin_thy}
  \dot{f}^{ }_\rmii{GW} (t,\vec{k})=\Gamma(k)\,\nB^{ }(k)=\frac{1}{8k}
  \int \! {\rm d}\Omega^{ }_{2\to2} \sum_{abc}
  \Bigl\vert\mathcal{M}^{ab}_{cG}
  (\vec{p}_1,\vec{p}_2;\vec{k}_1,\vec{k})\Bigr\vert^2 
  f^{ }_a(p^{ }_1)\,f^{ }_b(p^{ }_2)\,[1\pm f^{ }_c(k^{ }_1)]
  \;,
\end{equation}
where we have neglected $f^{ }_\rmii{GW} (t,\vec{k})$ on the right-hand side. 
The sum runs over all $abc\in$ SM (Standard Model) 
particle and antiparticle degrees of freedom  
and thus over all $ab\to c G$ processes,
with $G$ denoting the graviton.
$\vert\mathcal{M}^{ab}_{cG}(\vec{p}_1,\vec{p}_2;\vec{k}_1,\vec{k})\vert^2$ 
is the corresponding matrix element squared,
summed over all degeneracies of each species. 
For the SM in the symmetric phase, these are 
spin, polarization, colour, weak isospin and generation. 
For $k\sim \pi T$ the contribution of thermal masses is suppressed, 
so the external states can be considered massless (thermal masses
are only needed for the IR-divergent part 
of the squared amplitudes, cf.\ \se\ref{ss:htl}). 
The prefactor $1/8k$ is a combination of $1/2k$ from 
the phase space measure, 
$1/2$ for the graviton polarization degeneracy, and $1/2$ for 
the symmetry factor for identical initial state particles; 
in the cases where $a\ne b$ this factor
is compensated for by their being counted twice
in the sum over $abc$.
The thermal distributions 
$f_i$ correspond to $\nB^{ }$
and $\nF^{ }$ for bosons and fermions, respectively, with  
$[1\pm f_c(k_1)]$ implying $[1+ \nB^{ }(k_1)]$ in the former case 
and $[1- \nF^{ }(k_1)]$ in the latter.

The main challenge is the determination of 
the matrix elements squared, which requires the 
derivation of Feynman rules for all graviton-SM 
couplings and the computation of the tree-level
amplitudes. Given the large number of vertices and processes, 
and the associated opportunities for error, 
we have adopted automated techniques, originally developed for 
collider physics. We first used \textsc{FeynRules}~\cite{feynrules}, 
which can derive Feynman rules from a given Lagrangian.
We applied it to the  Lagrangian
describing the symmetric-phase SM coupled to gravitons, i.e.
\begin{equation}
  \label{sm_grav}
  \mathcal{L}^{ }_\rmii{SM+G}=\mathcal{L}^{ }_\rmii{SM}- 
  \frac{\sqrt{32 \pi}}{2 m_\rmii{Pl}}
  \, h_{\mu\nu}T^{\mu\nu}_\rmii{SM}
  \;,
\end{equation}
where the SM energy-momentum tensor $T^{\mu\nu}_\rmii{SM}$ contains 
also the trace part. The kinetic term for gravitons can be omitted, 
as they are external states in our computation.

Using the appropriate interface~\cite{feynrulestoFA},
\textsc{FeynRules} can generate a \textit{model file}
for \textsc{Feyn\-Arts}~\cite{FA}
(unfortunately, sometimes manual fixes of the generation
and SU(2) index assigments were needed).
This package and its companion \textsc{FormCalc}~\cite{FC} 
were then used to generate, evaluate and square all amplitudes, 
summing over the relevant degeneracies.\footnote{%
 We have also looked into several other packages, 
 however have not identified a procedure that would be simpler
 than the one described here. 
 }    
The handling of spin, vector boson polarization and colour is available 
in \textsc{FormCalc}, whereas SU(2) algebra and tensor boson polarization had
to be implemented. For the latter, we proceeded as follows. 
\textsc{FeynArts} assigns to external tensor bosons a polarization 
tensor $\epsilon^\lambda_{\mu\nu}(\vec{k})$ which is written, 
using a common factorization formula 
(cf.,\  e.g.,\ refs.~\cite{holstein1,holstein2}), as
\begin{equation}
  \label{poltensor}
  \epsilon^\lambda_{\mu\nu}(\vec{k})
  \; \equiv \; 
  \epsilon^\lambda_{\mu}(\vec{k})\,\epsilon^\lambda_{\nu}(\vec{k})
  \;,
\end{equation}
with $\epsilon^\lambda_{\mu}(\vec{k})$ the transverse polarization vector 
of a massless gauge boson. 
Upon taking $\vec{k}=k\,\vec{e}^{ }_z$ and the circular polarization vectors 
$\epsilon^\lambda_{\mu}(\vec{k})=1/\sqrt{2}(0,-1^\lambda,-i,0)$, 
$\lambda=1,2$, it is easy to verify
that the polarization sum satisfies
\begin{equation}
  \label{poltensorsum}\sum_\lambda 
   \epsilon^\lambda_{\mu\nu}(\vec{k})\,
  \epsilon^{\lambda\,*}_{\alpha\beta}(\vec{k})=
  L^{ }_{\mu\nu;\alpha\beta}
  \;,
\end{equation} 
with $L$ as defined in eq.~\eqref{defL}. 
We implemented this form of the tensor polarization 
sum as a \textsc{Mathematica} routine interfaced with 
the \textsc{Mathematica} output of \textsc{FeynArts}/\textsc{FormCalc}. 
The resulting matrix elements
have an apparent dependence on the projectors $\PT_{ }$, 
which again disappears by applying eq.~\eqref{magic}. 

Upon generating and evaluating all processes and plugging 
the results in eq.~\eqref{kin_thy}, we find
\begin{eqnarray}
 \hspace{-0.8cm} 
 \Gamma(k)\,\nB^{ }(k) \!\! &=& \!\! 
   \frac{1}{8k}
   \frac{32\pi}{m_\rmii{Pl}^2}
   \int \! {\rm d}\Omega^{ }_{2\to2} \bigg\{
   \nonumber\\
   \label{allbosons}
   &+&\nB^{ }(p_1)\,\nB^{ }(p_2)\,[1+\nB^{ }(k_1)] 
    \bigl(g_1^2+15g_2^2+48g_3^2\bigr)
   \left(\frac{s t}{u}+\frac{s u}{t}+\frac{t u}{s}\right)\\
   &-&\nF^{ }(p_1)\,\nB^{ }(p_2)\,[1-\nF^{ }(k_1)]
   \bigg[ 6 |h^{ }_t|^2 t 
   +\bigl(10g_1^2+18g_2^2+48g_3^2\bigr)
   \frac{s^2+u^2}{t}\bigg]\label{fbf}\\
   &-&\nB^{ }(p_1)\,\nF^{ }(p_2)\,[1-\nF^{ }(k_1)]
   \bigg[ 6 |h^{ }_t|^2 u 
   +\bigl(10g_1^2+18g_2^2+48g_3^2\bigr)
   \frac{s^2+t^2}{u}\bigg]\label{bff}\\
   &+&\nF^{ }(p_1)\,\nF^{ }(p_2)\,[1+\nB^{ }(k_1)]
  \bigg[ 6 |h^{ }_t|^2 s 
   +\bigl(10g_1^2+18g_2^2+48g_3^2\bigr)
   \frac{t^2+u^2}{s}\bigg]\label{ffb}\bigg\}
  \;. \hspace*{4mm} \la{boltzmann}
\end{eqnarray}
This expression agrees with the one obtained 
by plugging eqs.~\nr{cut_gg_2}--\nr{cut_fxg_2}
into eqs.~\eqref{C_Phi}, \eqref{Phi} and \eqref{Gamma_k}. 
To verify the agreement, relabellings 
$\vec p_1\leftrightarrow \vec p_2$ (and $t\leftrightarrow u$) 
as well as use of the identity
$
 \mathcal{N}^{ }_{\tau_1;\sigma_1\sigma_2}
 =   
        n^{ }_{\sigma_1} (p^{ }_1)
      \,n^{ }_{\sigma_2} (p^{ }_2) 
 \,[1 + n^{ }_{\tau_1}(k^{ }_1)]
 \,n^{-1}_{\tau_1\sigma_1\sigma_2}(p^{ }_1 + p^{ }_2 - k^{ }_1) 
$ 
are needed. 

In obtaining 
the fermionic parts of the total rate, 
i.e.\ \eqs\eqref{fbf}--\eqref{ffb},
we have not written out terms which arise from an odd number 
of $\gamma^{ }_5$ matrices in Dirac traces, 
since they vanish under the $\int \! {\rm d}\Omega^{ }_{2\to2}$ integration.
Specifically, these terms appear 
in the $fg\to fG$ processes and their crossings, 
with $f$ a fermion and $g$ a gauge boson. 

We also note
that the automated procedure fixes the gauge group factors, 
multiplicities and charge assignments to those specific
for the SM; the coefficients multiplying 
the coupling constants are not
obtained in terms of $\Nc$, $\nG$ and $\nS$. 
Focussing on sub-processes, it is easy to reinstate group theory factors. 
For instance, the $g_3^2$-part of \eq\nr{allbosons} corresponds to 
the matrix elements squared for the gluonic scattering $gg\to gG$, yielding
\begin{equation}
  \label{gggG}
  \left\vert\mathcal{M}^{gg}_{gG}
  (\vec{p}_1,\vec{p}_2;\vec{k}_1,\vec{k})\right\vert^2=
  \frac{32\pi}{m_\rmii{Pl}^2}\, 2 (\Nc^2-1)\Nc\, g_3^2
  \left(\frac{s t}{u}+\frac{s u}{t}+\frac{t u}{s}\right)
 \;. 
\end{equation}


Recently, there has been much work on factorizing
graviton amplitudes into photon amplitudes multiplied by 
kinematic factors, 
say $f\gamma\to f G$ versus $f\gamma\to f\gamma$
(cf.,\ e.g.,\ 
refs.~\cite{holstein1, holstein2} and references therein).
It is not clear to us, however, whether all the terms in 
\eqs\nr{allbosons}--\nr{boltzmann} could be related to 
photon production or scattering rates. 


We conclude this section by stressing that kinetic theory 
and its automated implementation are {\em not} sufficient for determining 
the leading-order gravitational wave production rate. 
Indeed, as discussed in \ses\ref{ss:ir} and \ref{ss:htl}, 
phase space integrals over matrix elements squared lead to 
IR divergences, 
related to soft gauge-boson exchange. The divergences
need to be subtracted and subsequently Hard Thermal Loop
resummed. An even more dramatic departure 
from the simple scattering picture is needed at smaller momenta, 
$k\sim \alpha_s^2 T$, where elementary particle states need to be 
replaced by hydrodynamic modes~\cite{qualitative}.

%
\subsection{Phase space integrals}
\la{ss:phase_space}

The next step is to carry out the phase space integral  
$
  \int \! {\rm d}\Omega^{ }_{2\to2} 
$
for the cuts in \eqs\nr{cut_gg_2}--\nr{cut_fxg_2}
or the matrix elements squared in \eqs\nr{allbosons}--\nr{boltzmann}.
For this task it is helpful to employ the parametrization
introduced in ref.~\cite{bb2}.\footnote{%
 If one is considering spectral functions off the light cone, 
 more complicated structures 
 $\sim \PT_{ } \mathcal{K}^4 / ( u t )$
 appear,  
 which require a refined parametrization if 
 a two-dimensional integral representation
 is desired~\cite{master,gj}. 
 } 
We discuss separately the treatment of $t$ and $s$-channel cases
($u$-channel can always be transformed into $t$-channel). 

%
\subsubsection{$t$-channel}

Consider the phase space integral
\ba
 \Gamma^{t}_{\tau_1;\sigma_1\sigma_2} & \equiv & 
  \int \! {\rm d}\Omega^{ }_{2\to2} \, 
 \mathcal{N}^{ }_{\tau_1;\sigma_1\sigma_2}
 \biggl\{ 
    a^{ }_1\, \frac{ s^2 + u^2 }{t}
  + a^{ }_2\, t 
 \biggr\} 
 \;. \la{t_channel} 
\ea
The idea is to insert 
$
 1 = \int \! {\rm d}^4 Q \, 
  \delta^{(4)}(\mathcal{P}^{ }_1 - \mathcal{K}^{ }_1 - \mathcal{Q})
$
in the integral. Then the energy-momentum conservation constraint
inside $ {\rm d}\Omega^{ }_{2\to2} $ can be written as 
$
 \delta^{(4)}(\mathcal{Q} + \mathcal{P}^{ }_2 - \mathcal{K})
$.
We can now integrate over $\vec{p}^{ }_2$ and $\vec{k}^{ }_1$
by using the spatial parts of the Dirac $\delta$'s, 
leaving $q^{ }_0,\vec{q}$ and $\vec{p}^{ }_1$ 
as the integration variables. The temporal Dirac $\delta$'s fix 
two angles as 
\be
 \vec{q}\cdot\vec{k} = \frac{q^2 - q_0^2 + 2 k q^{ }_0}{2}
 \;, \quad
 \vec{q}\cdot\vec{p}^{ }_1 = \frac{q^2 -q_0^2 + 2 p^{ }_1 q^{ }_0}{2} 
 \;, \la{angles_t}
\ee
whereas kinematic variables become
\ba
 && \hspace*{-1cm}
 t =  q_0^2 - q^2 \;, \quad
 u = 2 (\vec{k}\cdot\vec{p}^{ }_1 - k p^{ }_1) \;, \quad
 s = -t -u \;.
\ea
The azimuthal average of powers of $\vec{k}\cdot\vec{p}^{ }_1$ can be 
computed by parametrizing
\be
 \vec{q} = (0,0,q) \;, \quad
 \vec{k} = k\, (\sin\chi,0,\cos\chi) \;, \quad
 \vec{p}^{ }_1 = p^{ }_1\, (\sin\theta\cos\varphi,\sin\theta\sin\varphi,
 \cos\theta) \;, 
\ee
and integrating over $\varphi$. 
Denoting
$\langle ... \rangle \equiv 
  \frac{1}{2\pi}\int_0^{2\pi}\! {\rm d}\varphi \, (...) $,
this yields
\ba
 \bigl\langle \vec{k}\cdot\vec{p}^{ }_1 \bigr\rangle & = & 
 \frac{(\vec{q}\cdot\vec{k})\,(\vec{q}\cdot\vec{p}^{ }_1)}{q^2}
 \;, \la{kp11}
 \\ 
 \bigl\langle (\vec{k}\cdot\vec{p}^{ }_1)^2 \bigr\rangle & = & 
 \frac{1}{2} 
 \biggl[ 
  k^2 p_1^2 - \frac{p_1^2 (\vec{q}\cdot\vec{k})^2}{q^2}
            - \frac{k^2 (\vec{q}\cdot\vec{p}^{ }_1)^2}{q^2}
  + \frac{3 (\vec{q}\cdot\vec{k})^2 (\vec{q}\cdot\vec{p}^{ }_1)^2 
         }{q^4}
 \biggr] 
 \;. \hspace*{8mm} \la{kp13} 
\ea
The scalar products appearing here can be eliminated 
through \eq\nr{angles_t}. 
Finally, the phase space distributions
from \eq\nr{phasespace} can be cast in the form
\ba
 \mathcal{N}^{ }_{\tau_1;\sigma_1\sigma_2} 
 & = & 
 \bigl[
  1 + n^{ }_{\tau_1\sigma_1}(p^{ }_1 - k^{ }_1) 
    + n^{ }_{\sigma_2}(p^{ }_2) 
 \bigr]
 \bigl[
  n^{ }_{\tau_1}(k^{ }_1) - n^{ }_{\sigma_1}(p^{ }_1) 
 \bigr]
 \nn 
 & = & 
 \bigl[
  1 + n^{ }_{\tau_1\sigma_1}(q^{ }_0) 
    + n^{ }_{\sigma_2}(k - q^{ }_0) 
 \bigr]
 \bigl[
  n^{ }_{\tau_1}(p^{ }_1 - q^{ }_0) - n^{ }_{\sigma_1}(p^{ }_1) 
 \bigr]
 \;,
\ea
thereby factorizing the $p^{ }_1$-dependence. 

Denoting 
\be
 q^{ }_{\pm} \; \equiv \; \frac{q^{ }_0 \pm q}{2}
 \;, 
\ee
the integration range of $p^{ }_1$ can be established as $(\qp^{ },\infty)$.
The integration measure contains no powers of $p^{ }_1$, 
whereas azimuthal averages yield powers up to $p_1^2$.
The integral reads
\ba
 && \hspace*{-1cm}
 \int_{\qp^{ } }^{\infty} \! {\rm d} p^{ }_1 \, 
 \bigl( \beta^{ }_0 + \beta^{ }_1 p^{ }_1 + \beta^{ }_2 p_1^2  \bigr) 
 \bigl[ n^{ }_{\tau_1}(p^{ }_1 - q^{ }_0) - n^{ }_{\sigma_1}(p^{ }_1)\bigr]
 \nn 
 &  = &  
 \bigl( \beta^{ }_0 + \beta^{ }_1\, \qp^{ } + \beta^{ }_2\, \qp^2  \bigr) 
 \, L^{ }_1 
 + 
 \bigl( \beta^{ }_1 + 2 \beta^{ }_2\, \qp^{ } \bigr) 
 \, L^{ }_2 
 + 
 \bigl( 2 \beta^{ }_2    \bigr) 
 \, L^{ }_3 
 \;,
\ea
where 
\ba
 L^{ }_1 
 & \equiv &
 T \Bigl[
   \ln\Bigl( 1 - \sigma^{ }_1 e^{-\qp^{ }/ T} \Bigr) 
 - 
   \ln\Bigl( 1 - \tau^{ }_1\, e^{\qm^{ }/ T} \Bigr) 
   \Bigr] 
 \;, \\
 L^{ }_2 
 & \equiv &
 T^2 \Bigl[
      {\rm Li}^{ }_2 
      \Bigl( \tau^{ }_1\, e^{\qm^{ }/ T} \Bigr) 
  - 
      {\rm Li}^{ }_2 
      \Bigl( \sigma^{ }_1 e^{ - \qp^{ }/ T} \Bigr) 
   \Bigr] 
 \;, \\ 
 L^{ }_3 
 & \equiv &
 T^3 \Bigl[
      {\rm Li}^{ }_3 
      \Bigl( \tau^{ }_1\, e^{\qm^{ }/ T} \Bigr) 
  - 
      {\rm Li}^{ }_3 
      \Bigl( \sigma^{ }_1 e^{ - \qp^{ }/ T} \Bigr) 
   \Bigr] 
 \;. 
\ea
All in all this results in 
\ba
 \Gamma^{t}_{\tau_1;\sigma_1\sigma_2} & = & 
 \frac{1}{(4\pi)^3 k}
 \int_{-\infty}^k \! {\rm d} q^{ }_0 
 \int_{|q^{ }_0|}^{2k - q^{ }_0}
  \!\! {\rm d}q \, 
 \bigl[
  1 + n^{ }_{\tau_1\sigma_1}(q^{ }_0) 
    + n^{ }_{\sigma_2}(k - q^{ }_0) 
 \bigr] \, (q^2 - q_0^2)
 \nn & \times &
 \biggl\{
 \frac{a^{ }_1[q^2 - 3 (q^{ }_0 - 2k)^2]
 [12 L^{ }_3 + 6 q L^{ }_2 + q^2 L^{ }_1 ]}
 {6 q^4}
 - \biggl( a^{ }_2 + \frac{2 a^{ }_1}{3} \biggr) L^{ }_1
 \biggr\} 
 \;. \la{Gamma_t_full}
\ea

The integral in \eq\nr{Gamma_t_full} is logarithmically 
IR divergent at small $q^{ }_0,q$. For the different 
statistics the divergent parts read
\ba
 \Gamma^t_{+;++} |^{ }_\rmii{IR} 
 \; \equiv \; 
 - 2  \Gamma^t_{-;-+} |^{ }_\rmii{IR} 
 & \equiv & 
 \frac{1}{(4\pi)^3 k}
 \int_{-\infty}^k \! {\rm d} q^{ }_0 
 \int_{|q^{ }_0|}^{2k - q^{ }_0}
  \!\! {\rm d}q \, 
 \bigl[
  1 + \nB^{ }(q^{ }_0) 
    + \nB^{ }(k - q^{ }_0) 
 \bigr] 
 \nn & \times &  \biggl\{ 
 - \frac{4 a^{ }_1 q^{ }_0 (q^2 - q_0^2) k^2 \pi^2 T^2 }{q^4}
 \biggr\} 
 \;, \la{div_t1} \\ 
 \Gamma^t_{-;+-} |^{ }_\rmii{IR} 
 \; \equiv \; 
 - \Gamma^t_{+;--} |^{ }_\rmii{IR} 
 & \equiv & 
 \frac{1}{(4\pi)^3 k}
 \int_{-\infty}^k \! {\rm d} q^{ }_0 
 \int_{|q^{ }_0|}^{2k - q^{ }_0}
  \!\! {\rm d}q \, 
 \bigl[
  1 - \nF^{ }(q^{ }_0) 
    - \nF^{ }(k - q^{ }_0) 
 \bigr] 
 \nn & \times &  \biggl\{ 
  \frac{42 a^{ }_1  (q^2 - q_0^2) k^2 \zeta(3) T^3 }{q^4}
 \biggr\} 
 \;. \la{div_t4} \hspace*{5mm} 
\ea

%
\subsubsection{$s$-channel}

The $s$-channel phase space integral is defined as 
\ba
 \Gamma^{s}_{\tau_1;\sigma_1\sigma_2} & \equiv & 
  \int \! {\rm d}\Omega^{ }_{2\to2} \, 
 \mathcal{N}^{ }_{\tau_1;\sigma_1\sigma_2}
 \biggl\{ 
  b^{ }_1 \, \frac{t^2}{s} + b^{ }_2 \, s
 \biggr\} 
 \;. \la{s_channel} 
\ea
This time we insert 
$
 1 = \int \! {\rm d}^4 Q \, 
  \delta^{(4)}(\mathcal{P}^{ }_1 + \mathcal{P}^{ }_2 - \mathcal{Q})
$
in the integral, whereby the energy-momentum conservation constraint
inside $ {\rm d}\Omega^{ }_{2\to2} $ can be written as 
$
 \delta^{(4)}(\mathcal{Q} - \mathcal{K}^{ }_1 - \mathcal{K})
$.
We integrate over $\vec{p}^{ }_1$ and $\vec{k}^{ }_1$
by using the spatial parts of the Dirac $\delta$'s, 
leaving $q^{ }_0,\vec{q}$ and $\vec{p}^{ }_2$ 
as the integration variables. The temporal Dirac $\delta$'s fix 
two angles as 
\be
 \vec{q}\cdot\vec{k} = \frac{q^2 - q_0^2 + 2 k q^{ }_0}{2}
 \;, \quad
 \vec{q}\cdot\vec{p}^{ }_2 = \frac{q^2 - q_0^2 + 2 p^{ }_2 q^{ }_0}{2} 
 \;, \la{angles_s}
\ee
whereas kinematic variables become
\ba
 && \hspace*{-1cm}
 s =  q_0^2 - q^2 \;, \quad
 t = 2 (\vec{k}\cdot\vec{p}^{ }_2 - k p^{ }_2) \;, \quad
 u = -s - t \;.
\ea
The azimuthal average of powers of $\vec{k}\cdot\vec{p}^{ }_2$ can be 
computed like in \eqs\nr{kp11}--\nr{kp13}, exchanging 
$\vec{p}^{ }_1 \leftrightarrow \vec{p}^{ }_2$. 
The phase space distributions from \eq\nr{phasespace}
are now cast in the form
\ba
 \mathcal{N}^{ }_{\tau_1;\sigma_1\sigma_2} 
 & = & 
 \bigl[
  1 + n^{ }_{\sigma_1}(p^{ }_1) 
    + n^{ }_{\sigma_2}(p^{ }_2) 
 \bigr]
 \bigl[
  n^{ }_{\tau_1}(k^{ }_1) - n^{ }_{\sigma_1\sigma_2}(p^{ }_1 + p^{ }_2) 
 \bigr]
 \nn 
 & = & 
 \bigl[
  1 + n^{ }_{\sigma_1}(q^{ }_0 - p^{ }_2) 
    + n^{ }_{\sigma_2}(p^{ }_2) 
 \bigr]
 \bigl[
  n^{ }_{\tau_1}(q^{ }_0 - k) - n^{ }_{\sigma_1\sigma_2}(q^{ }_0) 
 \bigr]
 \;, 
\ea
factorizing the dependence on $p^{ }_2$.
The integration range of $p^{ }_2$ can be established as $(\qm^{ },\qp^{ })$,
and powers up to $p_2^2$ appear, 
whereby the general integral reads
\ba
 && \hspace*{-1.5cm} 
 \int_{\qm^{ } }^{\qp^{ } } \! {\rm d} p^{ }_2 \, 
 \bigl( \beta^{ }_0 + \beta^{ }_1\, p^{ }_2 + \beta^{ }_2\, p_2^2
 \bigr) 
 \bigl[
  1 + n^{ }_{\sigma_1}(q^{ }_0 - p^{ }_2) 
    + n^{ }_{\sigma_2}(p^{ }_2) 
 \bigr]
 \nn 
 & = & 
 \beta^{ }_0 q
 + \frac{\beta^{ }_1 q q^{ }_0}{2}
 + \frac{\beta^{ }_2 q (q^2 + 3 q_0^2)}{12}
 \nn 
 & - & 
 \bigl( \beta^{ }_0 + \beta^{ }_1\, \qp^{ } + \beta^{ }_2\, \qp^2
  \bigr) 
 \, L^{+}_1 
  - 
 \bigl( \beta^{ }_1 + 2 \beta^{ }_2\, \qp^{ }
 \bigr) 
 \, L^{+}_2
  - 
 \bigl( 2 \beta^{ }_2 
  \bigr) 
 \, L^{+}_3
 \nn 
 & + & 
 \bigl( \beta^{ }_0 + \beta^{ }_1\, \qm^{ } + \beta^{ }_2\, \qm^2
 \bigr) 
 \, L^{-}_1 
  + 
 \bigl( \beta^{ }_1 + 2 \beta^{ }_2\, \qm^{ }
  \bigr) 
 \,
 L^{-}_2 
  + 
 \bigl( 2 \beta^{ }_2
 \bigr) 
 \, L^{-}_3 
 \;, 
\ea
where
\ba
 L^{\pm}_1 & \equiv & 
 T \Bigl[
   \ln\Bigl( 1 - \sigma^{ }_1 e^{-\qmp^{ }/ T} \Bigr) 
 - 
   \ln\Bigl( 1 - \sigma^{ }_2\, e^{-\qpm^{ }/ T} \Bigr) 
   \Bigr] 
 \;, \\ 
 L^{\pm}_{2} & \equiv & 
 T^2 \Bigl[
      {\rm Li}^{ }_2 
      \Bigl( \sigma^{ }_2\, e^{- \qpm^{ }/ T} \Bigr) 
  + 
      {\rm Li}^{ }_2 
      \Bigl( \sigma^{ }_1 e^{ - \qmp^{ }/ T} \Bigr) 
   \Bigr] 
 \;, \\ 
 L^{\pm}_{3} & \equiv & 
 T^3 \Bigl[
      {\rm Li}^{ }_3 
      \Bigl( \sigma^{ }_2\, e^{ - \qpm^{ }/ T} \Bigr) 
  - 
      {\rm Li}^{ }_3 
      \Bigl( \sigma^{ }_1 e^{ - \qmp^{ }/ T} \Bigr) 
   \Bigr] 
 \;.  
\ea
All in all, this gives 
\ba
 \Gamma^{s}_{\tau_1;\sigma_1\sigma_2} & = & 
 \frac{1}{(4\pi)^3 k}
 \int_k^{\infty} \! {\rm d} q^{ }_0 
 \int_{|2k - q^{ }_0|}^{q^{ }_0}
  \!\! {\rm d}q \, 
 \bigl[
  n^{ }_{\tau_1}(q^{ }_0 - k) - n^{ }_{\sigma_1\sigma_2}(q^{ }_0) 
 \bigr] \, (q^2 - q_0^2)
 \nn & \times &
 \biggl\{
 \frac{b^{ }_1  [q^2 - 3 (q^{ }_0 - 2k)^2]
 [12(L^{-}_3 - L^{+}_3) - 6 q(L^{-}_2 + L^{+}_2) 
 + q^2(L^{-}_1 - L^{+}_1) ]  }
 {12 q^4}
 \nn 
 & & \; - 
 \frac{b^{ }_1 (q^{ }_0 - 2k)
 [2(L^{-}_2 - L^{+}_2) - q(L^{-}_1 + L^{+}_1) ]  }
 {2 q^2}
 - 
 \biggl( \frac{b^{ }_1}{3} + b^{ }_2 \biggr)
 (L^{-}_1 - L^{+}_1 + q)
 \biggr\} 
 \;. \nn \la{Gamma_s_full}
\ea

There is no IR divergence in the $s$-channel: 
would-be singular terms contain inverse powers of $q$, 
but the integration domain 
extends to small $q$ only around $q^{ }_0 = 2k$, where the 
integrand vanishes for all statistics
($\qpm^{ } = k + \rmO(q)$). 

%
\subsubsection{IR divergence}
\la{ss:ir}

Let us collect together the IR divergence
affecting the $2\leftrightarrow 2$ computation. 
Comparing \eqs\nr{cut_gg_2}--\nr{cut_fxg_2} with 
\eqs\nr{t_channel} and \nr{s_channel} we can extract
the coefficients appearing in \eqs\nr{div_t1} and \nr{div_t4}:
\ba
 \mathcal{C} \Phi^{ }_{g(g)} 
  \!\! & : & 
  a^{ }_1
 |^{ }_{+;++}
  = -4
 \;, \quad
  b^{ }_1
 |^{ }_{+;++}
  = -4
 \;, \hspace*{6mm} \la{cut_gg_3} \\[2mm] 
 \mathcal{C} 
    \Bigl[
       \Phi^{ }_{s(g)} 
     + \Phi^{ }_{g(s)} 
     + \Phi^{ }_{s|g} 
    \Bigr]
  \!\! & : & 
  a^{ }_1
 |^{ }_{+;++}
  = -1
 \;, \quad
  b^{ }_1
 |^{ }_{+;++}
  = -1
 \;, \hspace*{6mm} \la{cut_sxg_3} \\[2mm] 
 \mathcal{C}
    \Bigl[
       \Phi^{ }_{s(f)} 
     + \Phi^{ }_{f(s)} 
     + \Phi^{ }_{s|f} 
    \Bigr]
 \!\! & : & 
 a^{ }_2
 |^{ }_{-;-+}
  = 4 
 \;, \quad
 b^{ }_2
 |^{ }_{+;--}
  = 2 
 \;, \la{cut_sxf_3} \\[2mm] 
 \mathcal{C}
    \Bigl[
       \Phi^{ }_{f(g)} 
     + \Phi^{ }_{g(f)} 
     + \Phi^{ }_{f|g} 
    \Bigr]
 \!\! & : & 
 a^{ }_1
 |^{ }_{-;-+}
 = 4 
 \;, \quad
 b^{ }_1
 |^{ }_{+;--}
 = 4 
 \;. \la{cut_fxg_3}
\ea 
 The coefficient $a^{ }_1$ only comes with the statistical factors
 that were considered in \eq\nr{div_t1}, 
 so that the IR divergence shown
 in \eq\nr{div_t4} is absent. 
Adding prefactors according to \eq\nr{Phi} yields the total 
IR divergence of the $2\leftrightarrow 2$ contribution: 
\ba
 && \hspace*{-1.5cm} 
 \lim_{D\to 4}
 \im \bigl\{ 
  G^\rmii{R}_{12;12}
 \bigr\} 
 \bigr|^\rmii{IR}_{2\leftrightarrow 2} 
 \nn  
 & = & 
 \frac{1}{(4\pi)^3 k}
 \int_{-\infty}^k \! {\rm d} q^{ }_0 
 \int_{|q^{ }_0|}^{2k - q^{ }_0}
  \!\! {\rm d}q \, 
 \bigl[
  1 + \nB^{ }(q^{ }_0) 
    + \nB^{ }(k - q^{ }_0) 
 \bigr] 
 \frac{6 q^{ }_0 (q^2 - q_0^2) k^2\pi^2 T^2}{q^4}
 \nn & \times & 
 \biggl\{ 
   g_1^2 \, 
   \biggl[ \frac{\nS^{ }}{6} 
  +  \frac{\nG^{ }(\Nc^{ }+1)(\Nc^{ } + 2)}{12\Nc^{ }} \biggl] 
  \; + \;  
  3 g_2^2 \, 
  \biggl[ 
    \frac{2}{3} + \frac{\nS^{ }}{6} + \frac{\nG^{ }(\Nc^{ }+1)}{12}  
  \biggr] 
  \nn 
 & & \;  
 + \, \bigl( \Nc^{2} - 1 \bigr) g_3^2 \, 
  \biggl( \frac{\Nc^{ }}{3}+ \frac{\nG^{ }}{3} \biggr)  
 \biggr\} 
 \;. \la{IR_2to2}
\ea

%
\subsection{Hard Thermal Loop resummation}
\la{ss:htl}

The logarithmic IR divergence in \eq\nr{IR_2to2} 
can be eliminated through Hard Thermal
Loop resummation~\cite{htl3,htl4}. More precisely, as shown in ref.~\cite{bb2}
for a fermionic production rate and in ref.~\cite{qualitative} for the 
present observable, the infrared divergence is shielded through the
so-called Landau damping part of a resummed propagator, corresponding
physically to soft $t$-channel exchange.\footnote{%
 Originally this was shown in the context of photon production 
 in QCD~\cite{kls,rb,ar,amy2}. 
 } 
Thermal scatterings give an effective
mass to the exchanged gauge boson, whereby 
the logarithmic divergence turns into a finite logarithm, 
as we show in the remainder of this section.  
In principle there could be a similar contribution from 
soft $t$-channel fermion exchange, however in practice
there is no divergence at leading order, as we
demonstrate in appendix~\ref{sss:htl_fermion}. 
Scalar fields do not
experience Landau damping, so no discussion is needed for them. 
In the notation of \eq\nr{Phi}, we thus need to evaluate
\be
 G^\rmii{E}_{12;12} \bigr|^{ }_\rmii{HTL} =  
  \frac{2}{D(D-2)(D-3)}
  \biggl\{ 
  2 \nG^{ }(1 + \Nc^{ }) \Phi^{ }_f \bigr|^{ }_\rmii{HTL}
  + (2 + \Nc^{ }\CF^{ }) \Phi^{ }_g \bigr|^{ }_\rmii{HTL}
 \biggr\} 
 \;. \la{Phi_HTL}
\ee

Computing the diagram associated with 
$\Phi^{ }_g$ in \fig\ref{fig:graphs} with HTL-resummed propagators, 
the result reads\footnote{%
 The structure is the same for all three gauge groups, so we
 consider one of them as a representative.
 } 
\be
 \Phi^{ }_g \bigr|^{ }_\rmii{HTL} = 
 \frac{(D-2)L^{ }_{\mu\nu;\alpha\beta}}{2}\,
 \Tint{Q}
 4
   \Theta^{ }_{\alpha\beta;\rho\sigma}(Q,K-Q)
   \Delta^\rmii{HTL}_{\sigma\lambda}(K-Q)
   \Theta^{ }_{\mu\nu;\lambda\kappa}(K-Q,Q)
   \Delta^\rmii{HTL}_{\kappa\rho}(Q)
 \;, 
\ee
where $\Delta^\rmii{HTL}$ is the gauge propagator,  
\be
 \Delta^\rmii{HTL}_{\mu\nu}(K) = 
 \frac{ \PT_{\mu\nu} }{K^2 + \Pi^{ }_\rmii{T}(K)} +
 \frac{ \PE_{\mu\nu} }{K^2 + \Pi^{ }_\rmii{E}(K)} +
 \frac{\xi K^{ }_\mu K^{ }_\nu}{K^4}
 \;, 
\ee
with $\PT_{ }$ being the projector defined in \eq\nr{defL}, 
$\xi$ a gauge parameter, and
\be
 \PE_{\mu\nu} = \delta^{ }_{\mu\nu} - 
 \frac{K^{ }_\mu K^{ }_\nu}{K^4} - \PT_{\mu\nu}
 \;. 
\ee 
The tensor $\Theta$ parametrizes the cubic graviton-gauge vertex,   
\be
 \Theta^{ }_{\alpha\beta;\rho\sigma}(P,Q) \; \equiv \; 
 \bigl( P^{ }_\alpha \delta^{ }_{\mu\rho} 
      - P^{ }_\mu \delta^{ }_{\alpha\rho} \bigr)
 \bigl( Q^{ }_\beta \delta^{ }_{\mu\sigma} 
      - Q^{ }_\mu \delta^{ }_{\beta\sigma} \bigr) 
 \;.
\ee

The full HTL computation can be simplified by noting that 
in the diagrams of \fig\ref{fig:scat}, one of the gauge 
bosons attaching to the graviton vertex is always ``hard'' 
(i.e.\ with an external momentum $q \sim \pi T$)
and only one is ``soft'' (i.e.\ an internal $t$-channel rung).\footnote{%
 This is also the reason for why vertices do not need to be resummed. 
 } 
Adding to this that $\Theta$ projects out the longitudinal 
part of the propagator to which it is attached, permits us
to replace 
$
 \Delta^\rmii{HTL}_{\sigma\lambda}(K-Q) \to 
 2 \delta^{ }_{\sigma\lambda}/(K-Q)^2
$, 
where the factor 2 accounts for the two possibilities of picking
the hard line. Subsequently, after carrying out the contractions, we get
\ba
 \Phi^{ }_g \bigr|^{ }_\rmii{HTL} & \approx & \Tint{Q} \frac{4}{(K-Q)^2}
 \biggl\{ 
 \biggl( 
   \frac{1}{Q^2 + \Pi^{ }_\rmii{T}} - \frac{1}{Q^2 + \Pi^{ }_\rmii{E}}
 \biggr)
 \biggl[ 
    (D-3) \bigl[ \PT_\vec{q} \bigr]^2 
    \biggl( D - 2 - \frac{Q^2}{q^2} + \frac{D k^2 }{2 q^2} \biggr) 
 \nn 
  & - &
    \frac{D(D-3) Q^2 \PT_\vec{q}}{2} 
    \biggl( \frac{\vec{q}\cdot\vec{k}}{q^2} + \frac{Q^2}{4q^2} \biggr)
    + \frac{D(D-2)(D-3) Q^4}{8}
 \biggr]
 \nn 
 & + & 
 \frac{1}{Q^2 + \Pi^{ }_\rmii{E}}
 \biggl[ 
    (D-3)(D-2) \bigl[ \PT_\vec{q} \bigr]^2 
  - 
      \frac{D(D-3) Q^2 \PT_\vec{q}}{2}
    + \frac{D(D-2)(D-3) Q^4}{8}
 \biggr]
 \biggr\} 
 \;. \nn \la{Phi_g_HTL}
\ea
Furthermore, we may focus on the contribution that is largest
in the IR domain $q,q^{ }_0 \ll k$. This arises from 
the highest power of $k$ in the numerator, i.e.\ the term 
proportional to $k^2$ on the first line of \eq\nr{Phi_g_HTL}:
\be
 \Phi^{ }_g \bigr|^\rmii{IR}_\rmii{HTL} 
 \; \equiv \;
 \Tint{Q} \frac{2D(D-3)}{(K-Q)^2}
 \biggl( 
   \frac{1}{Q^2 + \Pi^{ }_\rmii{T}} - \frac{1}{Q^2 + \Pi^{ }_\rmii{E}}
 \biggr)\,
 \frac{k^2 \bigl[ \PT_\vec{q} \bigr]^2 }{q^2} 
 \;. \la{Phi_g_HTL_IR}
\ee 

At this point we write the Euclidean propagators in a spectral representation, 
\be
 \frac{1}{Q^2 + \Pi(Q)}
 = 
 \int_{-\infty}^{\infty} \! \frac{{\rm d}q^{ }_0} {\pi}
 \frac{\rho(q^{ }_0,q)}{q^{ }_0 - i q^{ }_n}
 \;, \quad
 \rho(q^{ }_0,q) \; \equiv \; 
 \im \biggl\{  \frac{1}{Q^2 + \Pi(Q)}
     \biggr\}^{ }_{q^{ }_n \to -i [q^{ }_0 + i 0^+]} 
 \;, \la{spectral}
\ee
carry out the Matsubara sum over $q^{ }_n$, and take the cut,
\ba
 \Gamma^{ }_\rmii{HTL} & \equiv & 
 \im\biggl\{ 
  \Tint{Q} \frac{1}{ (K-Q)^2[Q^2 + \Pi(Q) ]}
 \biggr\}^{ }_{k^{ }_n\to -i [k + i0^+]}
 \nn & = & 
 \int_{-\infty}^{\infty} \! {\rm d}q^{ }_0 \int_\vec{q} 
 \! \frac{\rho(q^{ }_0,q)}{2 \epsilon^{ }_{qk}}
 \Bigl\{ 
 \delta(q^{ }_0 - k - \epsilon^{ }_{qk})\, 
 \bigl[ \nB^{ }(\epsilon^{ }_{qk}) - \nB^{ }(q^{ }_0) \bigr]
 \nn  &   &
  + \,  
 \delta(q^{ }_0 - k + \epsilon^{ }_{qk})\, 
 \bigl[ 1 + \nB^{ }(q^{ }_0)  + \nB^{ }(\epsilon^{ }_{qk})\bigr]
 \Bigr\} 
 \;, \hspace*{4mm} \la{Gamma_HTL_def} 
\ea
where 
$
 \epsilon^{ }_{qk} \equiv |\vec{q-k}|
$.
Focussing on the soft contribution from the domain $q,q^{ }_0 \ll k$, 
only the latter channel gets kinematically realized. Carrying out the angular
integral, this contribution can be expressed as 
\be
 \Gamma^{ }_\rmii{HTL}
  \supset  
 \frac{1}{8\pi^2 k}
 \int_{-\infty}^{k} \! {\rm d}q^{ }_0 
 \int_{|q^{ }_0|}^{2k-q^{ }_0} \! {\rm d}q \, q \, 
 \bigl[ 1 + \nB^{ }(q^{ }_0)  + \nB^{ }(k - q^{ }_0) \bigr]
 \rho(q^{ }_0,q) \big|^{ }_{ 
 \vec{q}\cdot\vec{k} =  \frac{q^2 -q_\rmiii{0}^2 + 2 k q^{ }_\rmiii{0}}{2}}
 \;. \la{Gamma_HTL}
\ee

Inserting now the full structure of 
\eq\nr{Phi_g_HTL_IR} into \eq\nr{Gamma_HTL}, we get  
\ba
 \im \Bigl\{ 
  \Phi^{ }_g \bigr|^\rmii{IR}_\rmii{HTL}
 \Bigr\}^\rmi{ }_{k^{ }_n \to -i [k + i 0^+]} 
 & \stackrel{D\to 4}{\supset} &  
 \frac{1}{8\pi^2 k}
 \int_{-\infty}^{k} \! {\rm d}q^{ }_0 
 \int_{|q^{ }_0|}^{2k-q^{ }_0} \! {\rm d}q \, q\,  
 \bigl[ 1 + \nB^{ }(q^{ }_0)  + \nB^{ }(k - q^{ }_0) \bigr]
 \nn & \times & 
 \biggl\{ 
   \frac{8 k^2 \bigl[ \PT_\vec{q} \bigr]^2 }{q^2}
   \Bigl[ \rho_\rmii{T}^{ }(q^{ }_0,q) -
          \rho_\rmii{E}^{ }(q^{ }_0,q)
   \Bigr]
 \biggr\}^{ }_{ 
 \vec{q}\cdot\vec{k} =  \frac{q^2 -q_\rmiii{0}^2 + 2 k q^{ }_\rmiii{0}}{2}}
 \!\! . \hspace*{5mm} \la{Phi_g_HTL_IR2}
\ea
The angular constraint implies that
\be
 q_\perp^2 \; \equiv \; \PT_\vec{q} = (q^2 - q_0^2)
 \frac{(k - \qp^{ })(k-\qm^{ })}{k^2}
 \; \stackrel{q_{\pm}^{ }\ll k}{\approx} \; 
 q^2 - q_0^2
 \;. \la{qperp} 
\ee
The last step is invoked in order to carry out the resummation 
only for the leading term in an expansion in $q^{ }_0,q$, 
i.e.\ in the regime where there is an actual IR-divergence. 

We now apply \eq\nr{Phi_g_HTL_IR2} combined
with the insertion of \eq\nr{qperp} in two different ways. 
The first is to ``re-expand''
the result in the form of a weak-coupling expansion. In other words, the HTL 
spectral functions are evaluated for large $q,q^{ }_0$,
whereby they become
\be
 \rho^{ }_\rmii{T} \to \frac{\pi \mE^2 q^{ }_0}{4 q^3(q^2 - q_0^2)}
 \;, \quad
 \rho^{ }_\rmii{E} \to - \frac{\pi \mE^2 q^{ }_0}{2 q^3(q^2 - q_0^2)}
\;.
\ee
Here the Debye mass $\mE^{ }$ reads, 
in the case of the different gauge groups, 
\ba
 m_\rmii{E1}^2 & = &  g_1^2 T^2 \, 
 \biggl[ 
   \frac{\nS^{ }}{6} + 
   \frac{\nG^{ }(\Nc^{ }+1)(\Nc^{ }+2)}{12\Nc^{ }}
 \biggr]
 \;, \la{debye1} \\
 m_\rmii{E2}^2 & = &  g_2^2 T^2
 \biggl[
   \frac{2}{3} +  
   \frac{\nS^{ }}{6} + 
   \frac{\nG^{ }(\Nc^{ }+1)}{12}
 \biggr]
 \;, \la{debye2} \\ 
 m_\rmii{E3}^2 & = & g_3^2 T^2
 \biggl(
   \frac{\Nc^{ }}{3} + 
   \frac{\nG^{ }}{3} 
 \biggr)
 \;. \la{debye3}
\ea
In this way we find 
\ba
 \im \Bigl\{ 
  \Phi^{ }_g \bigr|^\rmii{IR}_\rmii{HTL}
 \Bigr\}^\rmi{expanded}_{k^{ }_n \to -i [k + i 0^+]} 
 & = &  
 \frac{1}{8\pi^2 k}
 \int_{-\infty}^{k} \! {\rm d}q^{ }_0 
 \int_{|q^{ }_0|}^{2k-q^{ }_0} \! {\rm d}q \,  
 \bigl[ 1 + \nB^{ }(q^{ }_0)  + \nB^{ }(k - q^{ }_0) \bigr]
 \, \Lambda(q^{ }_0,q)
 \nn & \times & 
   \frac{6 \pi q^{ }_0 (q^2 - q_0^2) k^2 \mE^2}{q^4}
 \;. \la{Phi_g_HTL_expanded}
\ea
Here a function $\Lambda$ has been introduced, with the property 
$\lim_{q^{ }_0,q\to 0}\Lambda = 1$. It can be chosen at will
outside of the domain where the resummation is implemented, 
given that its effects cancel up to higher-order corrections
(cf.\ the discussion below \eq\nr{resummation}).
 
Adding the prefactor from \eq\nr{Phi_HTL} and resolving the
different gauge groups, 
\be
 \frac{2 (2 + \Nc^{ }\CF^{ })\mE^2 }{D(D-2)(D-3)}  
 \; \stackrel{D\to 4}{\to} \;
 \frac{m_\rmii{E1}^2 + 3 m_\rmii{E2}^2 + (\Nc^2 - 1) m_\rmii{E3}^2 }{8}
 \;, \la{resolve}
\ee
we reproduce the IR divergence from \eq\nr{IR_2to2} in the domain
where $\Lambda = 1$. 

The second way is that we evaluate the HTL contribution as such. 
This could be computed numerically after inserting the 
full spectral functions $\rho^{ }_\rmii{T,E}$
into \eq\nr{Phi_g_HTL_IR2}, but through an opportune choice
of the weighting function $\Lambda$ it can also be
determined analytically, by making use of a sum rule~\cite{sum1,sum2}.
First, according to \eq\nr{qperp}, we can substitute 
$q^2 \approx q_0^2 + q_\perp^2$, and use then $q_\perp^{ }$ and $q^{ }_0$
as integration variables. 
Second, for $q^{ }_0 \ll T$, 
the Bose distribution $\nB^{ }(q^{ }_0) \approx T/q^{ }_0$
dominates over the terms $1 + \nB^{ }(k-q^{ }_0)$ that are of order unity. 
It is helpful to employ this simplification, which can 
be implemented by choosing $\Lambda = \Lambda^\star$, where
\be
 \bigl[ 1 + \nB^{ }(q^{ }_0)  + \nB^{ }(k - q^{ }_0) \bigr]
 \, \Lambda^\star(q^{ }_0,q)
 \; \equiv \; \frac{T}{q^{ }_0}
 \;. \la{Lambda}
\ee
We also note that the difference
$
 \rho_\rmii{T}^{ }(q^{ }_0,\sqrt{q_0^2 + q_\perp^2}) -
 \rho_\rmii{E}^{ }(q^{ }_0,\sqrt{q_0^2 + q_\perp^2})
$
decreases rapidly 
at large $|q^{ }_0|$, whereby the integration range over $q^{ }_0$ can be
extended to positive infinity. Therefore 
\ba
 \im \Bigl\{ 
  \Phi^{ }_g \bigr|^\rmii{IR}_\rmii{HTL}
 \Bigr\}^\rmi{full}_{k^{ }_n \to -i [k + i 0^+]} 
 & \stackrel{\;\Lambda = \Lambda^\star}{\approx} &  
 \frac{k T}{\pi^2}
 \int_{-\infty}^{\infty} \! \frac{ {\rm d}q^{ }_0 }{q^{ }_0} 
 \int_{0}^{2k} \! {\rm d}q^{ }_\perp \, q^{ }_\perp\,  
 \frac{ q_\perp^4 
   \bigl[ \rho_\rmii{T}^{ }(q^{ }_0,q) -
          \rho_\rmii{E}^{ }(q^{ }_0,q)
   \bigr] } {q^2}
 \nn 
 & \stackrel{\rmii{\cite{sum1,sum2}}}{=} & 
 \frac{k T}{\pi}
 \int_{0}^{2k} \! {\rm d}q^{ }_\perp \, q^{3}_\perp\,  
 \biggl(
   \frac{1}{q_\perp^2} - \frac{1}{q_\perp^2 + \mE^2} 
 \biggr) 
 \nn 
 & = & 
 \frac{k T \mE^2}{2 \pi}
 \ln\biggl( 1 + \frac{4 k^2}{\mE^2} \biggr) 
 \;. \la{Phi_g_HTL_full}
\ea
This logarithmically enhanced term corresponds to that 
determined in ref.~\cite{qualitative}. 

The full contribution of HTL resummation can now be obtained
by subtracting the term in \eq\nr{Phi_g_HTL_expanded} and adding
that in \eq\nr{Phi_g_HTL_full}, 
\be
 \Delta \im \Bigl\{ \Phi^{ }_g \bigr|^{ }_\rmii{HTL} \Bigr\} 
 \; \equiv \;  
 \im \Bigl\{ 
  \Phi^{ }_g \bigr|^\rmii{IR}_\rmii{HTL}
  \Bigr\}^\rmi{full}_{ }             
 - 
 \im \Bigl\{ 
  \Phi^{ }_g \bigr|^\rmii{IR}_\rmii{HTL}
  \Bigr\}^\rmi{expanded}_{ }             
 \;.  \la{resummation}
\ee
Given that for $q^{ }_0,q \gg \mE^{ }$ 
the full and expanded HTL spectral functions agree 
up to terms of $\rmO(g^4)$, the 
influence of $\Lambda$ 
drops out in this difference, however the same choice needs to 
be made in both terms (we chose $\Lambda = \Lambda^\star$).  
The gauge groups are resolved as in \eq\nr{resolve}. 
The subtraction term is evaluated together with \eq\nr{Gamma_t_full}, 
rendering the latter IR finite. 

%
\section{Numerical results}
\la{se:numerics}

\begin{figure}[t]

\hspace*{-0.1cm}
\centerline{%
 \epsfysize=7.5cm\epsfbox{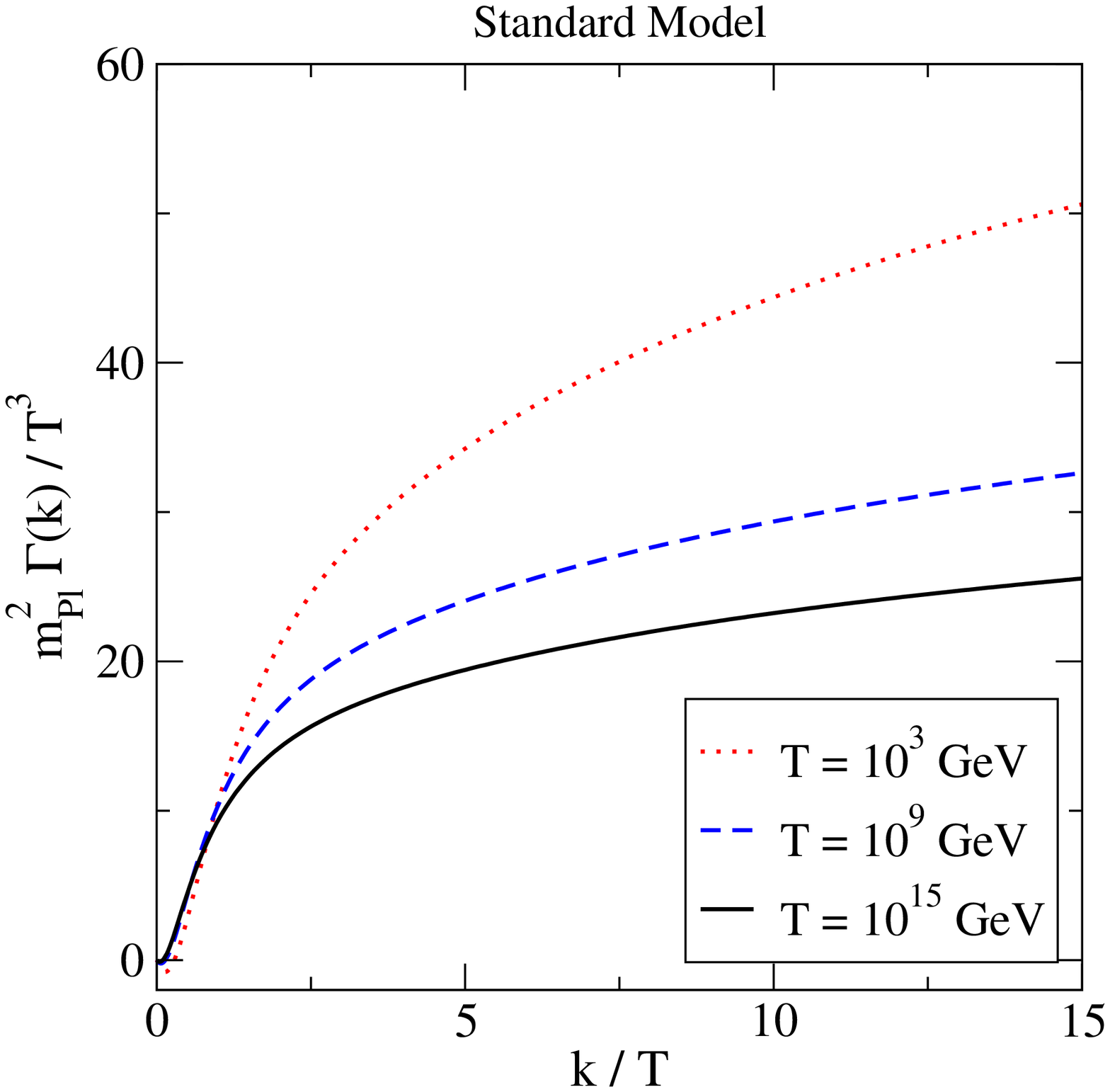}%
 \hspace{0.5cm}%
 \epsfysize=7.5cm\epsfbox{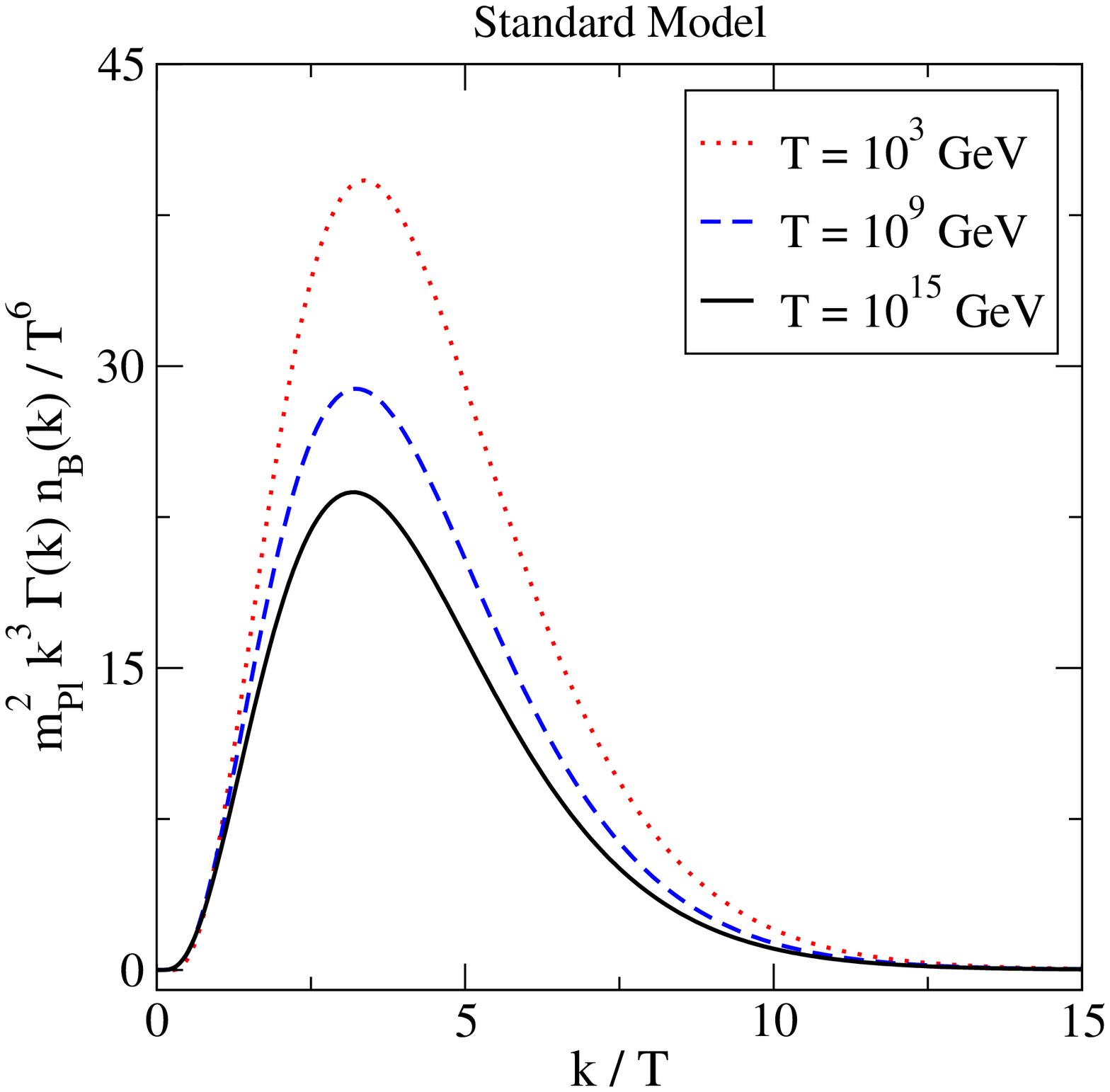}%
}

\caption[a]{\small
  Left: examples of 
  the interaction rate $\Gamma(k)$ from \eq\nr{Gamma_k}
  at a few representative temperatures, normalized 
  to $T^3/m_\rmii{Pl}^2$.
  The interaction rate decreases in these units with temperature, 
  because the most important running couplings become smaller. 
  Right: the combination 
  $m_\rmii{Pl}^2\, k^3\, \Gamma(k)\, \nB^{ }(k)/T^6$ 
  that plays a role for the production
  rate of the energy density carried by gravitational radiation. 
}

\la{fig:spectra}
\end{figure}

Inserting the integrals from \eqs\nr{Gamma_t_full} and \nr{Gamma_s_full}, 
with coefficients from \eqs\nr{cut_gg_3}--\nr{cut_fxg_3}, into \eq\nr{Phi}, 
and adding the resummation from \eq\nr{resummation}, we can 
determine the interaction rate $\Gamma(k)$ from \eq\nr{Gamma_k}. 
For the running couplings and Debye masses appearing in these
expressions, we use values specified in \se{4} of ref.~\cite{Debye}. 

In \fig\ref{fig:spectra}, 
$\Gamma(k)$ is plotted both as
$m_\rmii{Pl}^2\, \Gamma(k) /T^3$ 
and 
in the combination appearing in the energy density production rate, 
$m_\rmii{Pl}^2\, k^3\,\Gamma(k)\, \nB^{ }(k)/T^6$,   
at $T \approx 10^3, 10^9, 10^{15}$ GeV. In the units chosen, 
the rates decrease slowly with the temperature, due to the
running of $g_2^2$, $g_3^2$ and $h_t^2$. 

We remark that
$\Gamma(k)$ has a (barely visible) negative dip for $k/T \to 0$. 
In this region many of our approximations, 
taken under the assumption $k\sim \pi T$,
fail. Most importantly, HTL resummation with one hard 
and one soft gauge boson in 
$\Phi^{ }_g$, as described in sec.~\ref{ss:htl}, only works
correctly for $k\gg \mE^{ }$.\footnote{%
  For $k\gg \mEtiny^{ }$, we could actually replace
  the argument of the logarithm in \eq\nr{Phi_g_HTL_full} with just
  $4k^2/\mEtiny^2$, as the difference between these is
  parametrically of $\rmO(g^4)$.
  For $k \ll \mEtiny^{ }/2$, however, $\ln(1+4k^2/\mEtiny^2)$ 
  is small and positive,
  whereas $\ln(4k^2/\mEtiny^2)$ is large and negative. \label{foot:log} 
  That said, our result is formally incomplete for $k \lsim \mEtiny^{ }$, 
  as is practically any available thermal production rate as of today, 
  including that of photons from QCD.} 
This is neither new nor specific to graviton production:
previous calculations of gravitino \cite{gravitino1,gravitino2,gravitino3}, 
axion \cite{axion1,axion} and axino \cite{axino} production saw the same
issue. In fact, the negative dips were typically much larger
(cf.,\ e.g.,\ \fig{3} of~ref.\cite{axino}).
The reason for the difference can be traced back to 
the way in which HTL resummation was implemented in these works,
following ref.~\cite{braatenyuan}. 
Even if the method agrees with ours for $k \sim \pi T$
up to terms of $\rmO(g^4)$, it differs for $k\sim \mE^{ }$, 
in ways related to the discussion in footnote~\ref{foot:log}.
Remarkably, our implementation of HTL resummation 
avoids large negative dips 
without resorting to partial, gauge-dependent resummations 
of higher-order effects that were 
introduced in refs.~\cite{gravitino3} and \cite{axion} 
for gravitino and axion production, respectively.
These calculations could be revisited with our method, 
by finding the appropriate coefficients $a^{ }_i$ and $b^{ }_i$  
for eqs.~\nr{Gamma_t_full} and \nr{Gamma_s_full}, and taking over
our implementation of HTL resummation. 

%
\section{Cosmological implications}
\la{se:cosmo}

\begin{figure}[t]

\hspace*{-0.1cm}
\centerline{%
 \epsfysize=7.5cm\epsfbox{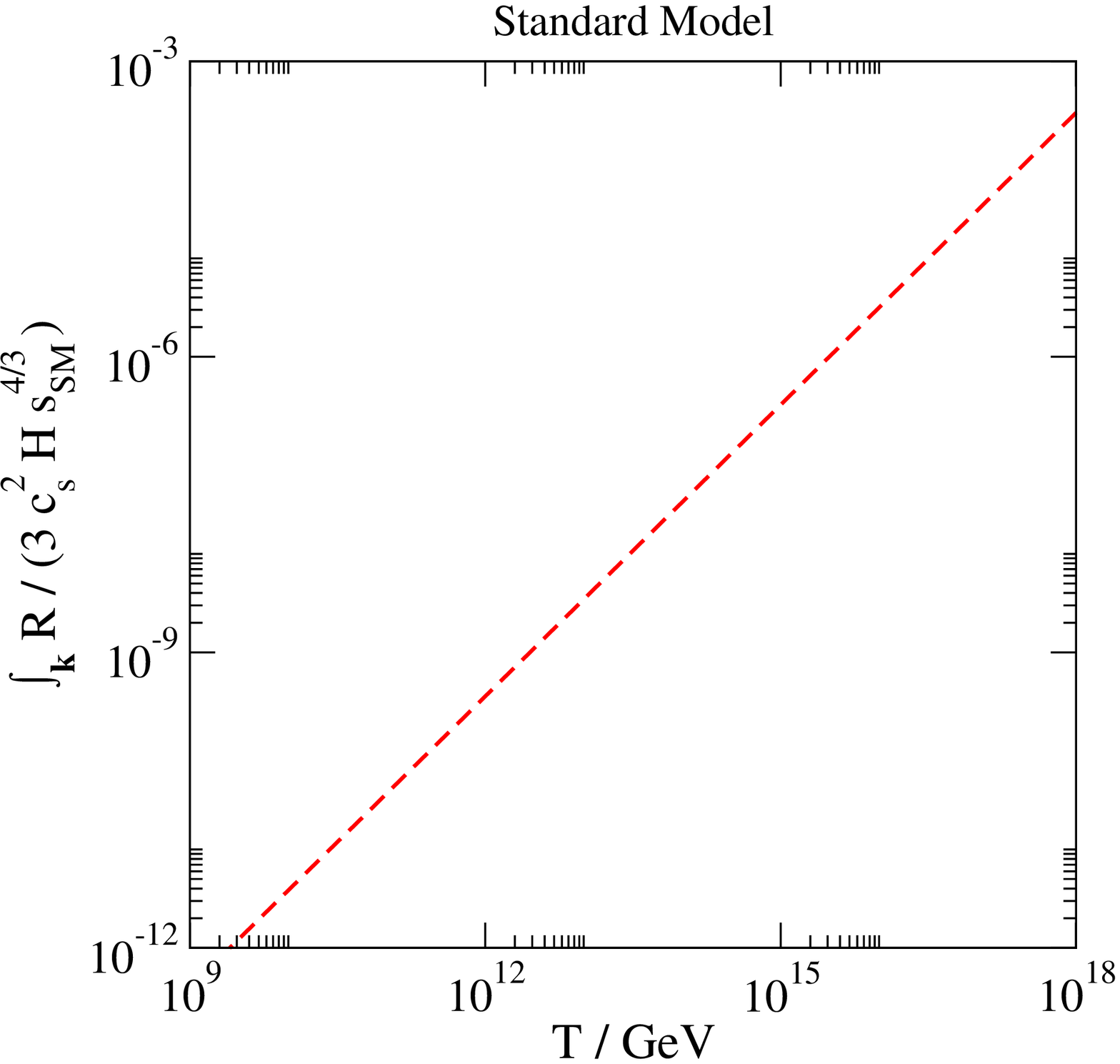}%
 \hspace{0.5cm}%
 \epsfysize=7.5cm\epsfbox{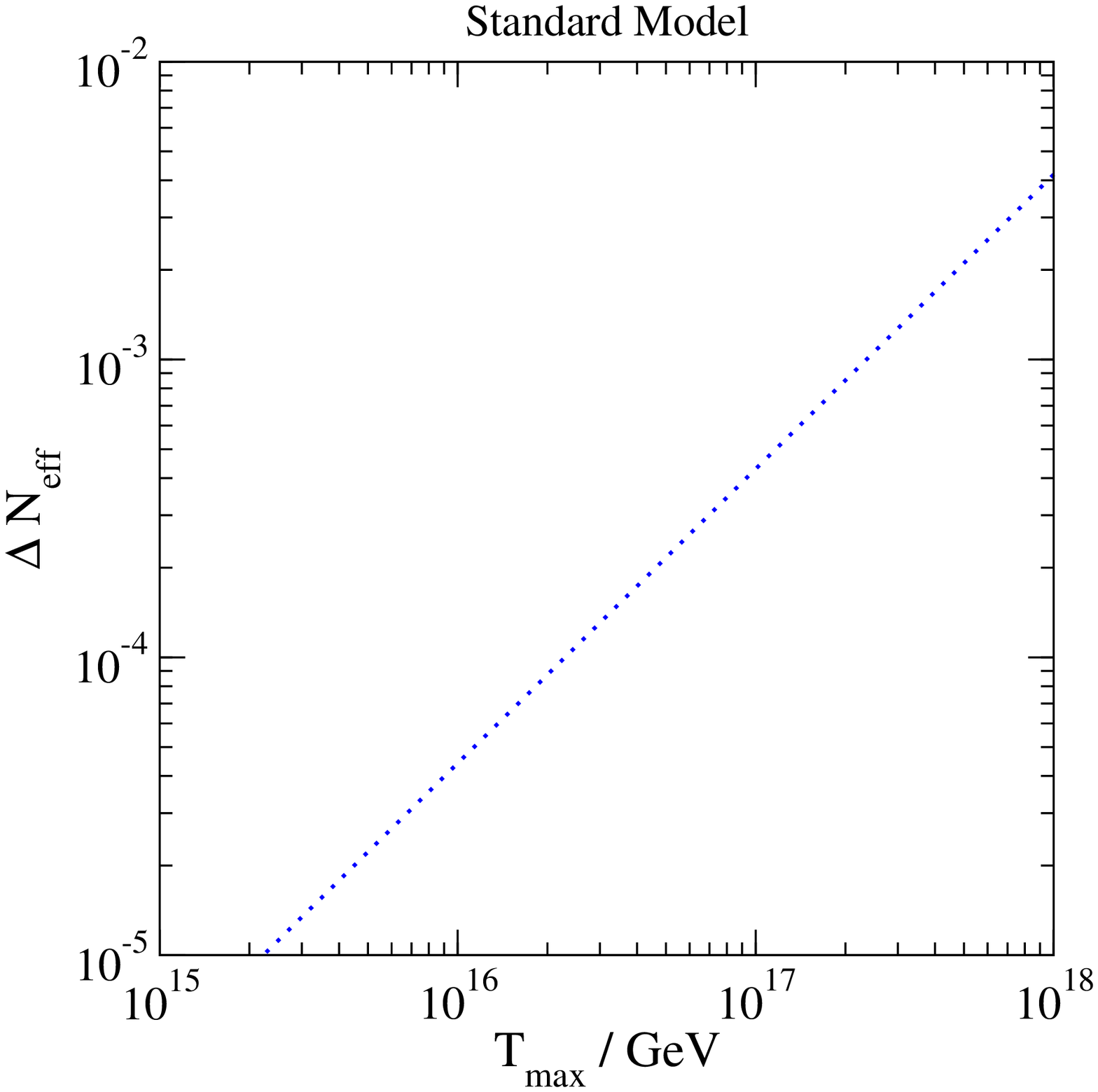}%
}

\caption[a]{\small
  Left: 
  the integrated production rate of the 
  energy density carried by gravitational radiation, 
  normalized as in \eq\nr{cos_int}, 
  as a function of the temperature. 
  Only the high-temperature end plays 
  a significant role. 
  Right: 
  the contribution of the gravitational energy density 
  to the parameter $N^{ }_\rmi{eff}$
  (cf.\ \eq\nr{Neff}), as a function of the 
  highest temperature of the radiation epoch. Once the 
  experimental determination of $N^{ }_\rmi{eff}$ reaches 
  the current theoretical precision, 
  $\Delta N^{ }_\rmi{eff} \sim 10^{-3}$, reheating
  temperatures above $T^{ }_\rmi{max}\approx 2\times 10^{17}$
  GeV can be constrained. 
}

\la{fig:Neff}
\end{figure}

As a final step we 
embed the production rate in an expanding cosmological 
background and compute 
\be
 \Delta N^{ }_\rmi{eff} 
 \; \equiv \; \frac{8}{7}\biggl(\frac{11}{4}\biggr)^{\fr43}
 \frac{e^{ }_\rmii{GW}(T^{ }_0)}{e^{ }_\gamma(T^{ }_0)}
 \;, \la{Neff}
\ee
where the final temperature 
can be chosen as $T^{ }_0 \sim 0.01$~MeV and 
$e^{ }_\gamma \equiv \pi^2 T_0^4/15$ is the energy density carried by photons.
The constraints originating from 
$N^{ }_\rmi{eff}$ are analogous in spirit to
the constraints on $e^{ }_\rmii{GW}$ considered
in refs.~\cite{Neff1,Neff2} (see also~\cite{rev}), and recently
$N^{ }_\rmi{eff}$ itself was invoked in ref.~\cite{hl}. 
The uncertainties of the Standard Model
prediction of $N^{ }_\rmi{eff}$ continue to be discussed in the literature
(cf., e.g., refs.~\cite{Neff3,Neff5,Neff4} and references therein), 
being around $\Delta N^{ }_\rmi{eff} \sim 10^{-3}$, 
whereas the  current experimental accuracy is
$\Delta N^{ }_\rmi{eff} \sim 10^{-1}$~\cite{planck},
which is expected to be reduced by an order of magnitude 
by future facilities~\cite{cmbs41}.
We consider the uncertainty of
the Standard Model prediction, $\Delta N^{ }_\rmi{eff} \sim 10^{-3}$,
to set an interesting sensitivity goal for considerations concerning
the gravitational background. 

Denoting by 
$
 H \equiv \sqrt{8\pi e^{ }_\rmii{SM}/(3 m_\rmi{Pl}^2)}
$
the Hubble rate, 
by 
$
 s^{ }_\rmii{SM}
$
the Standard Model entropy density, 
and by $c_s^2$ the speed of sound squared, 
the energy density at $T^{ }_0$ can be obtained as~\cite{qualitative} 
\be
 \frac{ e^{ }_\rmii{GW} (T^{ }_0) }{s^{4/3}_\rmii{SM}(T^{ }_0)}
 \; = \;
 \int_{0}^{\ln\bigl(\frac{ T^{ }_\rmii{max}}{ T^{ }_0 }\bigr)}
 \! \frac{{\rm d}x}{3 c_s^2 H}
 \frac{\int_\vec{k} R(T,k)}{s^{4/3}_\rmii{SM}(T)}
 \;, \quad
 x \; \equiv \; \ln \biggl( \frac{T^{ }_\rmii{max}}{T} \biggr)
 \;, \la{cos_int}
\ee
where the production rate $R$ is related to the damping 
coefficient $\Gamma$ from \eq\nr{rate_gen} through 
\be
 R(T,k) \; \equiv \; 2 k\, \Gamma(k)\, \nB^{ }(k)
 \;. \la{R_def}
\ee
The integrand of \eq\nr{cos_int} is illustrated 
in \fig\ref{fig:Neff}(left) as a function of the temperature. 
Clearly the integral
is dominated by the high-temperature end, 
so in practice we may restrict to temperatures above
the electroweak crossover, $T\sim 160$~GeV, for its determination.

The entropy dilution that takes place at low 
temperatures is accounted for by the factor 
$s^{4/3}_\rmii{SM}(T^{ }_0)$ in \eq\nr{cos_int}. 
We have adopted a prescription
for $s^{ }_\rmii{SM}$ which permits for its use even at $T < 2$~MeV
when neutrinos have decoupled (cf.\ the web page associated
with ref.~\cite{eos15} for the specification and for the 
numerical values that have been used\footnote{%
 The numerical values are attached to this publication 
 as supplementary material. 
 }).

Putting everything together, 
the contribution of the gravitational wave 
background to $N^{ }_\rmi{eff}$, obtained
from \eq\nr{Neff}, is shown in \fig\ref{fig:Neff}(right). 
Once the experimental accuracy reaches the level 
$\Delta N^{ }_\rmi{eff} \approx 10^{-3}$, maximal
temperatures above $2\times 10^{17}$~GeV can be constrained. 

%
\section{Conclusions and outlook}
\la{se:concl}

The main purpose of this paper has been to refine 
the estimate $T^{ }_\rmi{max} \lsim 10^{17...18}$~GeV that was obtained 
for the maximal temperature of the radiation epoch in ref.~\cite{qualitative}, 
by promoting the previous leading-logarithmic analysis to 
a full leading-order computation of the energy density
carried by gravitational radiation emitted by a Standard Model plasma. 
If the experimental determination of the parameter $N^{ }_\rmi{eff}$
can reach the current theoretical accuracy,
$\Delta N^{ }_\rmi{eff} \sim 10^{-3}$,
and no deviations from the Standard Model prediction are found, 
the refined estimate reads $T^{ }_\rmi{max} \le 2\times 10^{17}$~GeV. 
It is remarkable that this model-independent constraint 
is not much weaker than typical bounds on the reheating temperature
that are obtained by comparing model-dependent
inflationary predictions with 
Planck data~\cite{planck}. 

Most of the energy density carried 
by thermally produced gravitational radiation peaks
in the microwave frequency range today. 
Conceivably, this physics can be probed by tabletop experiments 
in the future~\cite{exp1,exp2,exp3,exp4,exp5,exp6,exp7,exp8,exp9}, 
even if the sensitivity goal is quite formidable. 

With future extensions in mind, 
we have displayed the technical steps of the computation in
quite some detail (cf.\ \se\ref{se:steps}). 
The partly automatized procedure to 
determine the matrix elements squared 
in \eqs\nr{allbosons}--\nr{boltzmann}
can be straightforwardly extended to other models. The IR subtraction
and thermal resummation that were described in \se\ref{ss:htl} must still
be adjusted accordingly, however we hope that our exposition lays
out these steps in a digestible fashion. Apart from 
graviton production in Beyond the Standard Model theories, 
this machinery can be applied to the production rates of
other particles coupling to a heat bath 
via non-renormalizable operators, such as 
gravitinos (with $M \ll \pi T$), axions and axinos.   
Indeed, as mentioned in sec.~\ref{se:numerics}, 
the phase space integration and resummation 
prescriptions of \ses\ref{ss:phase_space}, \ref{ss:htl}, 
which do not suffer from large, unphysical
negative contributions at small $k/T$, 
can be directly applied to the known matrix elements squared
in the literature~\cite{gravitino1,axino,axion1}.

%
\section*{Acknowledgements}

This work was partly supported by the Swiss National Science Foundation
(SNF) under grant 200020B-188712.

%
\appendix
\renewcommand{\thesection}{\Alph{section}} 
\renewcommand{\thesubsection}{\Alph{section}.\arabic{subsection}}
\renewcommand{\theequation}{\Alph{section}.\arabic{equation}}

%
\section{Soft $t$-channel fermion exchange}
\la{sss:htl_fermion}

We analyze in this appendix the fermion exchange part 
of \eq\nr{Phi_HTL}, {\it viz.}
$ 
 \Phi^{ }_f \bigr|^{ }_\rmii{HTL}
$, 
and show that no resummation is needed at leading order. 

Computing the diagram associated with 
$\Phi^{ }_f$ in \fig\ref{fig:graphs} within the HTL theory, 
the result reads\footnote{%
 The structure is the same for all fermions, so we
 consider one Dirac-like fermion as a representative.
 } 
\be
 \Phi^{ }_f \bigr|^{ }_\rmii{HTL} = 
 \frac{(D-2)L^{ }_{\mu\nu;\alpha\beta}}{2}\,
 \Tint{\{Q\}}
   \!\! \tr \bigl\{ 
   \Upsilon^{ }_{\mu\nu}(Q,K+Q)\,
   G^\rmii{HTL}_{ }(K+Q)\,
   \Upsilon^{ }_{\alpha\beta}(K+Q,Q)\,
   G^\rmii{HTL}_{ }(Q)
  \bigr\} 
 \;, 
\ee
where $G^\rmii{HTL}$ is the HTL-resummed fermion propagator,  
\be
 G^\rmii{HTL}_{ }(K) = 
 \frac{ i k^{ }_n \gamma^{ }_0 }{K^2 + \Pi^{ }_\rmii{W}(K)} +
 \frac{ i k^{ }_i \gamma^{ }_i }{K^2 + \Pi^{ }_\rmii{P}(K)} 
 \;, \la{G_prop}
\ee
and the tensor $\Upsilon$ parametrizes the cubic graviton-fermion vertex,   
\be
 \Upsilon^{ }_{\alpha\beta}(P,Q) \; \equiv \; 
 \frac{ 
         \gamma^{ }_\alpha \bigl( P^{ }_\beta + Q^{ }_\beta \bigr) 
       + \gamma^{ }_\beta \bigl( P^{ }_\alpha + Q^{ }_\alpha \bigr) 
 }{4} 
 \;.
\ee

Like in the gluonic case, we can replace 
one of the propagators by a free one
($\Pi^{ }_\rmii{W,P} \to 0$ in \eq\nr{G_prop}) and account for the 
associated symmetry by a factor 2. Taking the Dirac trace, 
this leads to 
\ba
 \Phi^{ }_f \bigr|^{ }_\rmii{HTL} & \approx & \Tint{\{Q\}} 
 \frac{2(D-3)}{(K+Q)^2}
 \biggl\{ 
 \frac{1}{Q^2 + \Pi^{ }_\rmii{W}} 
 \Bigl[ - D \PT_\vec{q} (q^{2}_n + q^{ }_n k^{ }_n) \Bigr]
 \nn  
 & + & 
 \frac{1}{Q^2 + \Pi^{ }_\rmii{P}} 
 \Bigl[ 
   4 \bigl[ \PT_\vec{q} \bigr]^2
 - D \PT_\vec{q} (q^2 + \vec{q}\cdot\vec{k})
 \Bigr]
 \biggr\}
 \;. \la{Phi_f_HTL}
\ea
Writing now 
\be
 q^{2}_n + q^{ }_n k^{ }_n = - (q^2 + \vec{q}\cdot\vec{k}) 
 + \frac{(K+Q)^2 + Q^2 - K^2}{2}
 \;, \la{qn}
\ee
and noting that $K^2$ vanishes on the light cone after analytic 
continuation and that $(K+Q)^2$ gives no cut as it cancels the 
free propagator, we can identify the most IR sensitive terms 
as those proportional to $\vec{q}\cdot\vec{k}$. 

Next, we invoke a spectral representation like in \eq\nr{spectral}, 
carry out the Matsubara sum over $q^{ }_n$, and take the cut,
\ba
 \widetilde \Gamma^{ }_\rmii{HTL} & \equiv & 
 \im\biggl\{ 
  \Tint{ \{ Q \} } \frac{1}{ (K+Q)^2[Q^2 + \Pi(Q) ]}
 \biggr\}^{ }_{k^{ }_n\to -i [k + i0^+]}
 \nn & = & 
  - \int_{-\infty}^{\infty} \! {\rm d}q^{ }_0 \int_\vec{q} 
 \! \frac{\rho(q^{ }_0,q)}{2 \tilde\epsilon^{ }_{qk}}
 \Bigl\{ 
 \delta(q^{ }_0 + k - \tilde\epsilon^{ }_{qk})\, 
 \bigl[ \nF^{ }(q^{ }_0) - \nF^{ }(\tilde\epsilon^{ }_{qk}) \bigr]
 \nn  &   &
  + \,  
 \delta(q^{ }_0 + k + \tilde\epsilon^{ }_{qk})\, 
 \bigl[ 1 - \nF^{ }(q^{ }_0)  - \nF^{ }(\tilde\epsilon^{ }_{qk})\bigr]
 \Bigr\} 
 \;, \hspace*{4mm} \la{Gamma_HTL_tilde_def}
\ea
where 
$
 \tilde\epsilon^{ }_{qk} \equiv |\vec{q+k}|
$.
Focussing on the soft contribution from the domain $q,q^{ }_0 \ll k$, 
only the first channel gives a contribution. It is convenient to 
substitute $q^{ }_0\to -q^{ }_0$ and make use of the antisymmetry
$
 \rho(- q^{ }_0,q) = - \rho(q^{ }_0,q)
$.
Carrying out the angular
integral, this yields 
\be
 \widetilde \Gamma^{ }_\rmii{HTL}
  \supset  
 \frac{1}{8\pi^2 k}
 \int_{-\infty}^{k} \! {\rm d}q^{ }_0 
 \int_{|q^{ }_0|}^{2k-q^{ }_0} \! {\rm d}q \, q \, 
 \bigl[ 1 - \nF^{ }(q^{ }_0)  - \nF^{ }(k - q^{ }_0) \bigr]
 \rho(q^{ }_0,q) \big|^{ }_{ 
 \vec{q}\cdot\vec{k} =  \frac{q_\rmiii{0}^2 - q^2 - 2 k q^{ }_\rmiii{0}}{2}}
 \;. \la{Gamma_HTL_tilde}
\ee

We note from the angular 
constraint in \eq\nr{Gamma_HTL_tilde} that for the most IR sensitive
contribution 
we can replace $\vec{q}\cdot\vec{k} \to - k q^{ }_0$. Combining this
with \eqs\nr{Phi_f_HTL} and \nr{qn} leads us to focus on  
\ba
 \im \Bigl\{ 
  \Phi^{ }_f \bigr|^\rmii{IR}_\rmii{HTL}
 \Bigr\}^\rmi{ }_{k^{ }_n \to -i [k + i 0^+]} 
 & \stackrel{D\to 4}{\equiv} &  
 \frac{1}{8\pi^2}
 \int_{-\infty}^{k} \! {\rm d}q^{ }_0 
 \int_{|q^{ }_0|}^{2k-q^{ }_0} \! {\rm d}q \, q\,  
 \bigl[ 1 - \nF^{ }(q^{ }_0)  - \nF^{ }(k - q^{ }_0) \bigr]
 \nn & \times &  
   8 q^{ }_0 \, \PT_\vec{q}\, 
   \bigl[ \rho_\rmii{P}^{ }(q^{ }_0,q) -
          \rho_\rmii{W}^{ }(q^{ }_0,q)
   \bigr]
 \;, \la{Phi_f_HTL_IR}
\ea
where $\PT_\vec{q}$ can be taken over from \eq\nr{qperp}. 

Again, we evaluate \eq\nr{Phi_f_HTL_IR} in two ways. Re-expanding in 
a strict weak-coupling expansion, the spectral functions become
\be
 \rho^{ }_\rmii{P} \to - \frac{\pi \mA^2 q^{ }_0}{4 q^3(q^2 - q_0^2)}
 \;, \quad
 \rho^{ }_\rmii{W} \to - \frac{\pi \mA^2 }{4 q q^{ }_0 (q^2 - q_0^2)}
 \;. \la{rho_f_asympt}
\ee
Here $\mA^{ }$ is a so-called asymptotic thermal mass~\cite{weldon}, 
which for quarks reads 
\be
  m^2_{q^{ }_\rmii{L}} = 
 \frac{(g_1^2 Y^2 + {3g_2^2}/{4} +  g_3^2 \CF^{ } )T^2}{4}
  \;, \quad
  m^2_{u^{ }_\rmii{R},d^{ }_\rmii{R}}
 = \frac{(g_1^2 Y^2 +  g_3^2 \CF^{ } )T^2}{4}
  \;, \la{mA}
\ee
where $Y$ denotes the hypercharge assignment as listed below 
\eq\nr{Dmu}. For the leptons, the SU(3) parts are absent. 
Inserting \eq\nr{rho_f_asympt} into \eq\nr{Phi_f_HTL_IR} yields
\ba
 \im \Bigl\{ 
  \Phi^{ }_f \bigr|^\rmii{IR}_\rmii{HTL}
 \Bigr\}^\rmi{expanded}_{k^{ }_n \to -i [k + i 0^+]} 
 & = &  
 \frac{1}{8\pi^2}
 \int_{-\infty}^{k} \! {\rm d}q^{ }_0 
 \int_{|q^{ }_0|}^{2k-q^{ }_0} \! {\rm d}q \,  
 \bigl[ 1 - \nF^{ }(q^{ }_0)  - \nF^{ }(k - q^{ }_0) \bigr]
 \nn & \times & 
   \frac{2 \pi (q^2 - q_0^2) 
   \mA^2}{q^2}
 \;. \la{Phi_f_HTL_expanded}
\ea
This is integrable (i.e.\ IR finite) 
at $q,q^{ }_0 \ll k$, and therefore does
{\em not} appear in \eq\nr{IR_2to2}. 

A complementary view on the soft fermion contribution can be 
obtained by evaluating \eq\nr{Phi_f_HTL_IR} like we did for 
the gauge contribution in \eq\nr{Phi_g_HTL_full}. Making use
of a sum rule derived in ref.~\cite{bb2}, and making a choice
analogous to \eq\nr{Lambda}, this gives
\ba
 \im \Bigl\{ 
  \Phi^{ }_f \bigr|^\rmii{IR}_\rmii{HTL}
 \Bigr\}^\rmi{full}_{k^{ }_n \to -i [k + i 0^+]} \hspace*{-3mm}
 \!\!& \approx &\!\!  
 \frac{1}{\pi^2}
 \int_{-\infty}^{\infty} \! {\rm d}q^{ }_0 \, q^{ }_0 
 \int_{0}^{2k} \! {\rm d}q^{ }_\perp \, q^{ }_\perp\,  
 \biggl[ \fr12 - \nF^{ }(k) \biggr]
 q_\perp^2 
   \bigl[ \rho_\rmii{P}^{ }(q^{ }_0,q) -
          \rho_\rmii{W}^{ }(q^{ }_0,q)
   \bigr] 
 \nn 
 \!\!& \stackrel{\rmii{\cite{bb2}}}{=} &\!\! 
 \frac{1}{2\pi}
 \biggl[ \fr12 - \nF^{ }(k) \biggr]
 \int_{0}^{2k} \! {\rm d}q^{ }_\perp \, q^{3}_\perp\,  
 \frac{\mA^2}{q_\perp^2 + \mA^2}  
 \;. \la{Phi_f_HTL_full}
\ea
The integral 
is dominated by $q_\perp^{ }\sim 2k$, yielding a contribution
of $\rmO(g^2 T^4)$ for $k\sim \pi T$. 
This is of leading order, but just a part
of the full result, not justifying any resummation. 

All in all, soft fermion exchange does {\em not} need to 
be resummed at leading order. 

%
\section{Magnitude of $1+n\leftrightarrow 2+n$ processes}
\la{ss:1to2}

%
\begin{figure}[t]
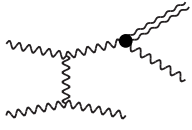


\vspace*{-1.0cm}%

\begin{eqnarray*}
 & &
 \hspace*{0cm}
 \GraphggC
\end{eqnarray*}

\vspace*{-1.2cm}%

\caption[a]{\small 
 An example of a 
 $2\leftrightarrow3$ scattering contributing 
 to gravitational wave production.
 The notation is as in \fig\ref{fig:scat}, 
 and the magnitude of these scatterings
 is estimated in appendix~\ref{ss:1to2}. 
} 
\la{fig:1nto2n}
\end{figure}
%

The processes we have considered in the main text, illustrated in 
\fig\ref{fig:scat}, correspond to $2\leftrightarrow 2$ scatterings. 
It may be asked if $1+n\leftrightarrow 2+n$ reactions also contribute. 
As Standard Model particles obtain thermal masses, whereas gravitons
remain massless, there is no phase space for such a process
at the Born level ($n=0$). However,
if one of the particles interacts before emitting a gravitational
wave ($n\ge 1$), so that it is set slightly off-shell, 
this argument no longer applies.
An example of this type of a ``bremsstrahlung'' process 
is shown in \fig\ref{fig:1nto2n}.
In the context of producing
photons or massless fermions from 
a thermal plasma, such processes do contribute at the same order as 
$2\leftrightarrow 2$ scatterings, and have to be summed
to all orders ($\sum_{n=0}^{\infty}$), through a procedure known as 
Landau-Pomeranchuk-Migdal (LPM) resummation~\cite{lpm1,amy1,bb1}. 
In footnote~1 of ref.~\cite{axion}, it has however been pointed out that 
such reactions are of subleading order for gravitational wave production. 
The purpose of this appendix is to confirm the assertion of ref.~\cite{axion},
which we do by employing light-cone variables similar to those normally
adopted for LPM resummation. 

In the notation of \eq\nr{Phi}, i.e.\ treating the gauge groups
on equal footing for a moment, 
the LPM contribution reads
\be
 G^\rmii{E}_{12;12} \bigr|^{ }_\rmii{LPM} =  
  \frac{2}{D(D-2)(D-3)}
  \biggl\{ 
  \nS^{ } \, \Phi^{ }_s \bigr|^{ }_\rmii{LPM}
  + 2 \nG^{ }(1 + \Nc^{ }) \Phi^{ }_f \bigr|^{ }_\rmii{LPM}
  + (2 + \Nc^{ }\CF^{ }) \Phi^{ }_g \bigr|^{ }_\rmii{LPM}
 \biggr\} 
 \;. \la{Phi_LPM}
\ee
In order to determine the three terms, we start by writing
their (vanishing) Born limits in a suggestive 
form.\footnote{%
 More precisely, to extract 
 the information, all terms contributing to the ``slope''
 towards the vanishing limit need to be included, which
 in the current context amounts to terms $\propto \mathcal{K}^4$.
 } 
According to \eqs\nr{Phi_s}--\nr{Phi_g}, 
the cuts read 
\ba
 \lim_{D\to 4} \im \bigl\{ \Phi^{ }_s   \bigr\} 
 \bigr|^{ }_{k_n\to -i[\omega+i0^+]} 
 & = & 
 4 \im \bigl\{ J^2_{11} \bigr\} 
 \bigr|^{ }_{k_n\to -i[\omega+i0^+]} 
 \;, \la{Phi_s_LPM} \\ 
 \lim_{D\to 4} \im \bigl\{ \Phi^{ }_f   \bigr\} 
 \bigr|^{ }_{k_n\to -i[\omega+i0^+]} 
 & = &  
 -2 \im \bigl\{ 2 \tilde J^2_{11} + \tilde J^1_{11} \bigr\} 
 \bigr|^{ }_{k_n\to -i[\omega+i0^+]} 
 \;, \hspace*{5mm} \\
 \lim_{D\to 4} \im \bigl\{ \Phi^{ }_g   \bigr\} 
 \bigr|^{ }_{k_n\to -i[\omega+i0^+]} 
 & = &  
 2 \im \bigl\{ 2 J^2_{11} + 4 J^1_{11} + J^0_{11} \bigr\} 
 \bigr|^{ }_{k_n\to -i[\omega+i0^+]} 
 \;, \la{Phi_g_LPM}
\ea
where the masters $J,\tilde J$ were defined in \eq\nr{J_def}
and we have kept $\omega\neq k$. 
Let us approach the light cone from above, setting
$\omega \equiv \sqrt{k^2 + M^2} $ with $M^2 \to 0^+$.
Adopting results from \eqs\nr{Gamma_HTL_def} and \nr{Gamma_HTL_tilde_def} 
and setting $\mathcal{Q}\to -\mathcal{Q}$ in the latter, we can write
\ba
 \Gamma^{ }_\rmii{LPM} 
 \!\! & \equiv & \!\!  
  \im\biggl\{ 
  \Tint{Q} 
  \frac{
   \alpha^{ }_{0} \,\bigl[ \PT_\vec{q} \bigr]^2
  + \alpha^{ }_1  \,\PT_\vec{q}\, K^2
  + \alpha^{ }_2\, K^4 
  }{ (K-Q)^2 Q^2}
  \biggr\}^{ }_{k^{ }_n\to -i [\omega + i0^+]}
  \nn & = & 
  \int_{-\infty}^{\infty} \! {\rm d}q^{ }_0 \int_\vec{q} 
 \! \frac{\rho^{ }_\rmi{free}(q^{ }_0,q)}{2 \epsilon^{ }_{qk}}
 \, 
 \bigl\{ 
    \alpha^{ }_{0} \,\bigl[ \PT_\vec{q} \bigr]^2
  - \alpha^{ }_1  \,\PT_\vec{q}\, M^2
  + \alpha^{ }_2\, M^4 
 \bigr\} 
   \la{Gamma_LPM_def}  \\  
 & \times & \!\!  
 \Bigl\{ 
 \delta\bigl( q^{ }_0 - \omega - \epsilon^{ }_{qk} \bigr)\, 
 \bigl[ n_\sigma^{ }(\epsilon^{ }_{qk}) - n_\sigma^{ }(q^{ }_0) \bigr]
 + 
 \delta\bigl( q^{ }_0 - \omega + \epsilon^{ }_{qk} \bigr)\, 
 \bigl[ 1 + n_\sigma^{ }(q^{ }_0)  + n_\sigma^{ }(\epsilon^{ }_{qk})\bigr]
 \Bigr\} 
 \;, \nonumber
\ea
where $\sigma = \pm$ takes care of statistics
according to \eq\nr{n_sigma}. The free spectral function reads
\be
 \rho^{ }_\rmi{free}(q^{ }_0,q) = \frac{\pi
 \bigl[
  \delta(q^{ }_0 - q ) - \delta(q^{ }_0 + q) 
 \bigr]
 }{2q}
 \;. \la{rho_free}
\ee
For $M^2 > 0$ the contribution comes from 
the second kinematic channel in \eq\nr{Gamma_LPM_def} 
combined with the first term in \eq\nr{rho_free}.

We now go over to light-cone coordinates, 
$
 \vec{q} = q^{ }_{\parallel}\, \vec{e}^{ }_\vec{k} + \vec{q}^{ }_\perp
$, 
so that 
\be
 \epsilon^{ }_{qk} = \sqrt{(k-q^{ }_\parallel)^2 + q_\perp^2}
 \;, \quad
 \PT_\vec{q} = q_\perp^2 
 \;.
\ee
The constraint $\delta(q^{ }_0 - q)$ is eliminated by integrating
over $q^{ }_\parallel$, which sets 
$
 q^{ }_\parallel = \sqrt{q_0^2 - q_\perp^2} 
$
(here we anticipate the overall sign to be positive,  
$q^{ }_\parallel \sim q^{ }_0 \in (0,k)$, cf.\ \eq\nr{Gamma_LPM}).
The remaining constraint 
$\delta(q^{ }_0 - \omega + \epsilon^{ }_{qk})$
implies  
\be
 M^2 
 \; = \; \omega^2 - k^2 
 \; = 2 \biggl[ \, 
 q_0^2 - k \sqrt{q_0^2 - q_\perp^2}
 + q^{ }_0 \sqrt{k^2 + q_0^2 - 2 k \sqrt{q_0^2 - q_\perp^2}}
 \, \biggr]
 \;.  
\ee
This can be expanded in $q_\perp^2/q_0^2$ and $q_\perp^2 / (k-q^{ }_0)^2$, 
assuming again $0 < q^{ }_0 < k$ to fix signs. Keeping
contributions up to $q_\perp^4$ in 
$
    \alpha^{ }_{0} \, q_\perp^4
  - \alpha^{ }_1 \, q_\perp^2 \, M^2
  + \alpha^{ }_2\, M^4 
$ 
and contributions up to $q_\perp^2$ inside 
$
 \delta(q^{ }_0 - \omega + \epsilon^{ }_{qk})
$, 
we find
\ba 
 \lim_{D\to 4} \im \bigl\{ \Phi^{ }_i  \bigr\} 
 \bigr|^{ }_{k_n\to -i[\omega +i0^+]} 
 & \stackrel{M^2 \,\approx\, 0\;\;\;}{=} &    
  \int_{-\infty}^{\infty} \!\!\! {\rm d}q^{ }_0 
  \, \kappa^{ }_i (q^{ }_0) \,
 \bigl[ 1 + n_\sigma^{ }(q^{ }_0)  + n_\sigma^{ }( k - q^{ }_0)\bigr]
 \nn 
 & \times & 
 \int_{\vec{q}^{ }_\perp} \!\! q_\perp^4 \, 
 \delta\biggl( - \frac{M^2}{2k} 
   + \frac{q_\perp^2}{2(k - q^{ }_0 )}
   + \frac{q_\perp^2}{2 q^{ }_0 } \biggr)
 \;. \la{Gamma_LPM}
\ea
It is clear from here that 
for $M^2 > 0$ the contribution originates from $0 < q^{ }_0 < k$.
However,  
we have removed the specifier $M^2\to 0^+$, because \eq\nr{Gamma_LPM} turns
out to be applicable for $M^2\to 0^-$ as well, with the contribution
originating from $q^{ }_0 < 0$ and $q^{ }_0 > k$ in that case. 

When the coefficients $\alpha^{ }_{0},\alpha^{ }_1,\alpha^{ }_2$ are inserted
into the prefactor according to \eqs\nr{Phi_s_LPM}--\nr{Phi_g_LPM}, 
the functions $\kappa^{ }_i$ in \eq\nr{Gamma_LPM} become
\be
 \kappa^{ }_s (q^{ }_0)  =  
 \frac{1}{2 q^{ }_0(k-q^{ }_0)}
 \;, \quad 
 \kappa^{ }_f (q^{ }_0)  =  
 \frac{q_0^2 + (k-q^{ }_0)^2}{4 q^{2}_0(k-q^{ }_0)^2}
 \;, \quad 
 \kappa^{ }_g (q^{ }_0)  =  
 \frac{q_0^4 + (k-q^{ }_0)^4}{4 q^{3}_0(k-q^{ }_0)^3}
 \;. \la{kappa_g} 
\ee
Up to overall conventions, 
$ \kappa^{ }_s $ and
$ \kappa^{ }_f $
agree with the prefactors cited for 
scalars and fermions in ref.~\cite{amy1}. 
The factor $ \kappa^{ }_g $ is similar to 
the prefactor for the gluon contribution to gluon emission
that was discussed in ref.~\cite{amy3}, however 
it is {\em not} exactly the same:
the latter has an additional $k^4$ in the numerator, guaranteeing
a symmetry between the three gluons involved. 

Let us now estimate the magnitude of the $1+n\leftrightarrow 2+n$
contributions. For this, we can set the virtuality to be 
parametrically $M^2 \sim g^2 T^2$, as it is at this scale
that thermal masses and scatterings of the type in 
\fig\ref{fig:1nto2n} play a role if 
$q^{ }_0 \sim k \sim \pi T$. Then \eq\nr{Gamma_LPM} implies that
$
 q_\perp^2 = M^2 q^{ }_0 (k - q^{ }_0)/k^2
$
and, 
up to logarithms in the case of $\kappa^{ }_g$, 
$\im\bigl\{ \Phi^{ }_i \bigr\} \sim M^4 \sim g^4T^4$.
This is suppressed by $\rmO(g^2)$ compared with 
the effects that we are interested in. 

\small{ 
%

}

\end{document}